\documentclass[superscriptaddress,showkeys,preprintnumbers,amsmath,amssymb,floatfix,prd]{revtex4}

\usepackage{graphicx}
\usepackage{dcolumn}
\usepackage{bm}
\usepackage{hyperref}
\usepackage{orcidlink}
\hypersetup{colorlinks,citecolor=blue}

\begin{document}

\preprint{}

\title{Study on deflection angle, shadow, quasinormal modes, and greybody factor of the black hole surrounded by quintessence in Rastall gravity}

\author{Susmita Sarkar \orcidlink{0009-0007-1179-2495}}
\email{susmita.mathju@gmail.com}
\affiliation{Department of Applied Science and Humanities, Haldia Institute of Technology, Haldia-721606, Purba Mednipur, West Bengal, India}

\author{Nayan Sarkar \orcidlink{0000-0002-3489-6509}}
 \email{  nayan.mathju@gmail.com}
\affiliation {Department of Mathematics, Karimpur Pannadevi College, Karimpur-741152, Nadia, West Bengal, India}

\author{Tuhina Manna \orcidlink{0000-0001-6171-7572}}
\email{ tuhinamanna03@gmail.com}
\affiliation{Department of Mathematics, 
St. Xavier’s College(Autonomous), Kolkata-700016,West Bengal, India }

\author{Moumita Sarkar}
\email{moumitasarkar1594@gmail.com}
\affiliation{Department of Mathematics, Jadavpur University, Kolkata-700 032, West Bengal, India}

\author{Farook Rahaman \orcidlink{0000-0003-0594-4783}}
\email{rahaman@iucaa.ernet.in}
\affiliation{Department of Mathematics, Jadavpur University, Kolkata-700 032, West Bengal, India}

\date{\today }

\begin{abstract}
 The present study focuses on investigating the deflection angle in the weak-field approximation, shadow, quasinormal modes using Lyapunov exponents, and lower bound of the greybody factor for a charged black hole surrounded by a quintessence field in Rastall gravity. The weak deflection angles are calculated using the Gauss-Bonnet method. They decrease with increasing impact parameter $b$ and charge $Q$, but gradually increase with increasing black hole mass $m$. Notably, the presence of a surrounding quintessence field in Rastall gravity leads to a higher deflection angle compared to Schwarzschild or Reissner–Nordström black holes with positive $\mathcal{N}_q$. The photon sphere and shadow of the black hole are analysed concerning the charge $Q$ and mass $m$; they shrink as
$Q$ increases and expand with increasing $m$. We further analyse the quasinormal modes of the black hole, explicitly derive the coordinate time Lyapunov exponent $\lambda_c$ and the quasinormal frequency $\omega$. In the eikonal limit, the Lyapunov exponent ensures that the real and imaginary parts of the quasinormal modes can be expressed by the frequency and instability time scale of the unstable null circular geodesics. Additionally, we derive the lower bounds of the greybody factor $\mathcal{G}_b$, it decreases for increasing charge $Q$ while the increasing mass $m$ enhances it. Importantly, all the findings reduce to those of the Reissner–Nordström black hole for $\mathcal{N}_q = 0$ and to the Schwarzchild black hole for $\mathcal{N}_q = Q = 0$. 

\end{abstract}

\maketitle

\section{Introduction}
In 1972 Peter Rastall \cite{Rastall1} proposed a modified gravity theory where the $T^\nu _{\mu;\nu}$=0 condition of covariant energy-momentum conservation, was violated and instead the divergence of $T_{\mu \nu}$ is considered proportional to the gradient of the Ricci scalar, i.e., 
\begin{equation}\label{eq:Rastall}
T^\nu _{\mu;\nu}=\lambda R^{,\nu}.
\end{equation}
So we have $T^\nu _{\mu;\nu}=0$ when $\lambda=0$ and it reduces to Einstein's equation in empty spacetime but differs from them in the presence of matter. However, Bianchi's identities on the Einstein tensor are unaltered.
It was claimed that Rastall gravity approach introduced a deep non-minimal coupling between gravity and matter and provides a phenomenological framework for identifying quantum effects in gravitational systems, specifically the breach of classical conservation laws, which is also observed in the $f(R, T)$ and $f(R, L_m)$ theories, where $R$, $T$, and $L_m$ denote the Ricci scalar, the trace of the energy-momentum tensor, and the matter Lagrangian, respectively. Additionally, the condition $T^u _{\mu;u} \neq 0 $ is empirically validated by the particle creation process in cosmology. It should be noted that while the equation given in Eq. (\ref{eq:Rastall}) was initially introduced by Rastall, it is not the only possible form. Any other form can be taken, provided $T^\nu_\mu$ becomes zero in flat spacetime \cite{Rastall2}. Thus, any general form can be proposed as follows:
\begin{equation}
T^\nu_{\mu;\nu}= A^\nu_\mu;_\nu,
\end{equation}
where $A_{\mu\nu}=A_{\nu\mu}$ and $A_{\mu\nu}$ and its derivatives are sufficiently small in flat spacetimes, and hence may be neglected. Thus, the key difference from Einstein gravity is that Rastall gravity considers a non-divergence-free energy-momentum.
In addition to Rastall's foundational work on Rastall gravity, numerous significant papers have emerged over the years.
Before comparing this modified gravity theory with general relativity it is worth noting that, Visser \cite{Visser} demonstrated Rastall gravity is equivalent to Einstein's gravity, whereas Darabi et.al.\cite{add1} argued to the contrary that these two theories are not equivalent and that, Rastall theory of gravity is more "open" to accepting the challenges of observational cosmology and quantum gravity. The difference between Rastall and Einstein gravity has been a subject of longstanding research \cite{add1.5}. Rastall gravity seems to be consistent with observational data on the age of the Universe, on the Hubble parameter \cite{add2} and with results obtained from helium nucleosynthesis \cite{add3}. Theoretically, it does not contradict the results obtained from general relativity in the matter-dominated era \cite{add4}, including the solar system tests \cite{add5} and solutions of anisotropic compact stars \cite{add6}. It is also claimed that Rastall gravity does not suffer from the entropy and age problems of the standard cosmological model \cite{add7}.  Inspired by this, our study aims to explore these claims by examining the various tests of gravity, namely the calculation of deflection angle to study the weak gravitational lensing, shadow, quasinormal modes using Lyapunov exponents, and greybody bound of a charged black hole surrounded by quintessence field in Rastall gravity. Since these tests are crucial for comparing any modified theory of gravity with Einstein gravity, we intend to analyze how the observational data from Einstein gravity aligns or contrasts with that from Rastall gravity. We hope this research will further prompt scholars to investigate the connection between the two theories.  In the last few years, significant research on the black hole shadow image, quasinormal modes and gravitational lensing has emerged in the quintessence field, greybody factors in Rastall gravity \cite{new1,new2,new3,new4,new5,new6,new7,new8,new9}.
Also, major research in the last decade was conducted by  \cite{yh17,Mor,Bron1,HeyMord,Spall,Morad1,Kumar,Ma} and \cite{Oliv}. Additionally, notable works include: \cite{Santos,Bez,Lobo1,Licata,Car} and \cite{Sal}, among others. It should be noted that Batista et al. have suggested that Rastall gravity could be viewed as an implementation of certain quantum effects in a curved spacetime \cite{Batista}. In cosmology, the appeal of Rastall gravity has been reflected in numerous publications, such as: \cite{cos1, Fabris,Bron,Darabi,Yuan,Fab,BatistaFab} and \cite{Morad}.\par
  This article deals with gravitational lensing in the weak field limit, known as weak lensing, using the Gauss-Bonnet theorem first proposed by Gibbons and Werner in \cite{gibbons1}. In 1916, Einstein anticipated the phenonmenon of gravitational lensing and gravitational waves as basic consequences of general relativity \cite{Einstein}. In 2015, this prediction was verified through the gravitational waves detected by the Laser Interferometer Gravitational-wave Observatory (LIGO) \cite{Abbot}. In general, weak gravitational lensing is characterized by the formation of a distorted image of the source into an arc like shape, smeared around the lens centre: an effect known as "shear". This is in contrast to the larger deflection suffered in the strong lensing case, when multiple separate images are observed. The third type of lensing is microlensing wherein a relatively smaller lens, ranging from planetary to stellar masses, is involved, and the microlensing event is itself a transient astronomical event, since the source, lens and observer all have relative proper motions, and their alignment to create an effective deflection of light is temporary; observed only as an apparent brightening of the source.\cite{Wam} The Gauss Bonnet method of obtaining deflection angle was later extended by Werner \cite{Werner2012} considering the Kerr-Randers optical geometry to find the angle of deflection of light by a Kerr black hole. Ishihara et al. \cite{ Ishihara2016,  Ishihara2017} and Ono et al. \cite{Ono2018, Ono2019} made significant contributions to this taking into account the corrections for finite distance between the source and observer. This was enhanced further by considering a non-trivial spacetime topology to compute the deflection angle of light caused by cosmic strings or a global monopole, as well as for rotating and non-rotating wormholes \cite{Jusufi2018a, Jusufi2018b, Jusufi2018c, Jusufi2018d,manna2023}.\par
Normal modes are inherent to closed systems, whereas quasinormal modes (QNMs) are associated with open systems. When a black hole undergoes perturbation, the immediate burst of radiation is succeeded by the ringdown phase, during which QNMs emerge as a result of the dissipative oscillation of spacetime. The QNMs depend entirely on the intrinsic parameters of a black hole and the specific type of perturbation that triggers them. As the black hole oscillates, it dissipates energy by emitting gravitational waves (GWs). These QNMs are described by complex frequencies: the real part indicates the frequency of the emitted GWs, while the imaginary part represents the damping. Hence, these frequencies represent the decay rates of the GWs.  
QNMs were first pointed out by Vishveshwara \cite{Vish} in gravitational wave calculations of the Schwarzschild black hole, while Press \cite{Press}
coined the term "quasi-normal" frequencies. Gogoi et al. \cite{dj23} investigated the thermodynamics, Joule‐Thomson expansion and optical behaviour of a Reissner‐Nordström‐anti‐de Sitter black hole in Rastall gravity surrounded by a quintessence field. Furthermore, the quasinormal modes, greybody factors, and absorption cross section of a de Sitter Reissner–Nordström black hole surrounded by a quintessence field in Rastall gravity were studied in Ref. \cite{dj24}. The study of the stability of black holes under small perturbations was initiated a long time back in the late 50s by Regge and Wheeler \cite{RG} and later by Zerilli \cite{Zer}.
The perturbations of relativistic stars in GR were first studied in the late 60s by Kip Thorne and his collaborators \cite{Kip1,Kip2,Kip3,Kip4}. Studies with respect to Schwarschild \cite{SQNM} and Kerr black holes\cite{KerrQNM1,KerrQNM2} including the first and second-order results in perturbation theory for head-on collisions of black holes \cite{Pullin} have also been done extensively. Binary black holes exhibit unstable orbits when they are in close proximity. Lyapunov exponents quantify the instability of orbits, holding observational significance as the phase of gravitational waves can become incoherent within a Lyapunov time, regardless of whether the orbits are regular or chaotic. Consequently, analyzing QNMs is a powerful tool in studying black hole geometry.

One of the main motivations behind using quintessence for this paper is to solve the most basic mystery in cosmology: the composition of the Universe. Data from cosmic microwave background (CMB) anisotropy spectrum\cite{mot1} and type 1A supernovae \cite{mot2,mot2.5} strongly support a spatially flat Universe undergoing  accelerated expansion.This results in a shortfall in the total energy density of the Universe, which can be explained by hypothesizing a time-varying, spatially inhomogeneous, negative pressure component of the cosmic fluid called quintessence \cite{mot3,mot4,mot5}. This approach may also be able to solve the "coincidence problem". Although cosmological constant is also taken to be a possible explanation \cite{mot6,mot6.5}, yet it requires extreme fine tuning to explain the expansion history of the Universe, since the value of $\omega$, the ratio of pressure($p$) to energy density($\rho$) is precisely -1 at all times. The idea of quintessence scalar field  on the other hand, is more dynamic, since then, $-1<\omega \leq 0$. Also the notion of "tracker models" \cite{mot7} facilitates a wide range of initial conditions to evolve into matching current matter density profile \cite{mot7.5,mot7.6}. Further there is a greater scope of quintessence field coupling or interacting with different forms of energy or gravity \cite{mot8}.

The paper is divided as follows: Section \ref{sec2} gives a brief discussion on charged black hole surrounded by the quintessence field in Rastall gravity; section \ref{sec3} develops a conceptual idea of the deflection angle using the Gauss-Bonnet method, while next in section \ref{sec4} the deflection angles are investigated for two different values of the parameters $\kappa \lambda$. Further in section \ref{sec5}, the deflection angles obtained using the two different values of the parameters $\kappa \lambda$ are compared with respect to the impact parameter, black hole charge and mass. Following this, in section \ref{sec6} the shadow cast by the black hole is explored, detailing the photon sphere and shadow radius; quasinormal modes via Lyapunov exponents are studied in section \ref{sec7}; while section \ref{sec8} deals with understanding the bounds on the greybody factor. Finally, our results and conclusions are discussed in section \ref{sec9}.  

\begin{figure}[!htbp]
\begin{center}
\begin{tabular}{rl}
\includegraphics[width=8cm]{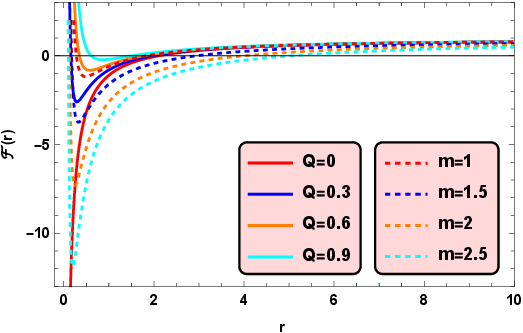}
\end{tabular}
\begin{tabular}{rl}
\includegraphics[width=8cm]{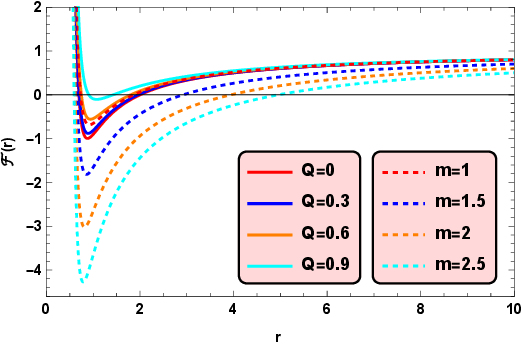}
\end{tabular}
\end{center}
\caption{Profiles of the metric function $\mathcal{F}(r)$ against the radial coordinate $r$ corresponding to  Left:  $\kappa\lambda = -3/2$, $\mathcal{N}_q = 0.1$ with $m = 1$ (Solid Line) and $Q = 0.5$ (Dashed Line); Right:  $\kappa\lambda = 3/2$ , $\mathcal{N}_q = -0.1$ with $m = 1$ (Solid Line) and $Q = 0.5$ (Dashed Line).}\label{fig1}
\end{figure}

\section{Charged black hole surrounded by quintessence in Rastall gravity }\label{sec2}

 Here, we are going to present the charged black hole solution surrounded by the quintessence field in Rastall's theory of gravity \cite{yh17}. For a spacetime with a Ricci scalar $R$ and an energy-momentum tensor $T_{\mu\nu}$ the Rastall’s hypothesis \cite{Rastall1, Rastall2} states that 
\begin{equation}
    T^{\mu\nu}_{~~~;\mu} = \lambda R^{,\nu},
\end{equation}

where $\lambda$ is the Rastall parameter that quantifies the deviation from the standard conservation law in general relativity. Therefore, the Rastall field equations can be expressed as
\begin{eqnarray}
    G_{\mu\nu}+\kappa\lambda g_{\mu\nu}R = \kappa T_{\mu\nu},
\end{eqnarray}
where $\kappa$ is the Rastall gravitational coupling constant. It is important to note that the above field equations reduce to the standard general relativity field equations when $\lambda = 0$ and $\kappa = 8\pi G_N$ with $G_N$ as the Newton gravitational coupling constant. Now, for the derivation of black hole solutions, we take into account  the general spherically symmetric spacetime metric, given by
\begin{equation}
    ds^2 = -\mathcal{F}(r)dt^2+\frac{1}{\mathcal{F}(r)}dr^2+r^2(d\theta^2+\sin\theta d\phi^2).\label{ds}
\end{equation}
where $\mathcal{F}(r)$ is the metric function that depends solely on the radial coordinate $r$. 

For the above metric (\ref{ds}), the non-zero components of the Rastall tensor defined as $H_{\mu\nu} = G_{\mu\nu}  + \kappa\lambda g_{\mu\nu}R$ are obtained as
\begin{eqnarray}
    H^0_{~0} &=& G^0_{~0}+\kappa\lambda R = -\frac{1}{\mathcal{F}}G_{00}+\kappa\lambda R =\frac{1}{r^2}\left(r\mathcal{F}'-1+\mathcal{F}\right)+\kappa\lambda R,\label{h1}
    \\
     H^1_{~1} &=& G^1_{~1}+\kappa\lambda R = \mathcal{F}G_{11}+\kappa\lambda R =\frac{1}{r^2}\left(r\mathcal{F}'-1+\mathcal{F}\right)+\kappa\lambda R,\label{h2}
      \\
     H^2_{~2} &=& G^2_{~2}+\kappa\lambda R = \frac{1}{r^2}G_{22}+\kappa\lambda R =\frac{1}{r^2}\left(r\mathcal{F}'+\frac{r^2}{2}\mathcal{F}''\right)+\kappa\lambda R,\label{h3}
      \\
     H^3_{~3} &=& G^3_{~3}+\kappa\lambda R = \frac{1}{r^2\sin^2\theta}G_{33}+\kappa\lambda R =\frac{1}{r^2}\left(r\mathcal{F}'+\frac{r^2}{2}\mathcal{F}''\right)+\kappa\lambda R,\label{h4}
\end{eqnarray}

with the Ricci scalar
\begin{eqnarray}
    R = -\frac{1}{r^2}\left(r^2 \mathcal{F}''+4r\mathcal{F}'-2+2\mathcal{F}\right).
\end{eqnarray}

Here, the "prime" sign stands for the derivative with respect to the radial coordinate $r$. Now, considering the non-zero components of the Rastall tensor $H^\mu_{~~\nu}$, the total energy-momentum tensor can be expressed as
\begin{eqnarray}
    T^\mu_\nu = \text{diag}\left(T^0_{~~0},~ T^1_{~~1}, ~T^2_{~~2}, ~T^3_{~~3}\right),
\end{eqnarray}

\begin{figure}[!htbp]
\begin{center}
\begin{tabular}{rl}
\includegraphics[width=8.3cm]{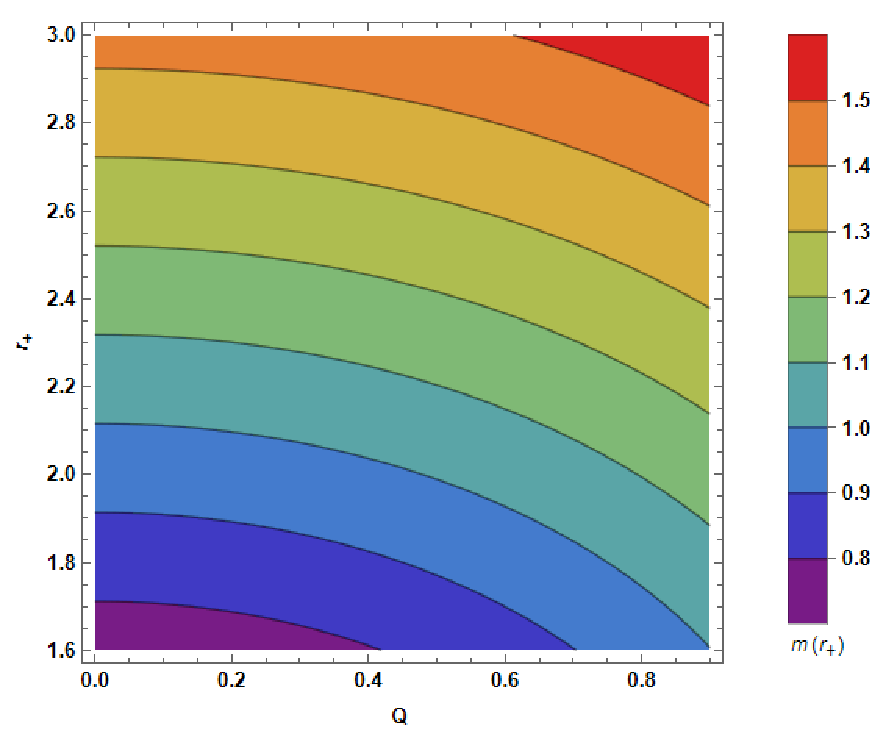}
\end{tabular}
\begin{tabular}{rl}
\includegraphics[width=8.3cm]{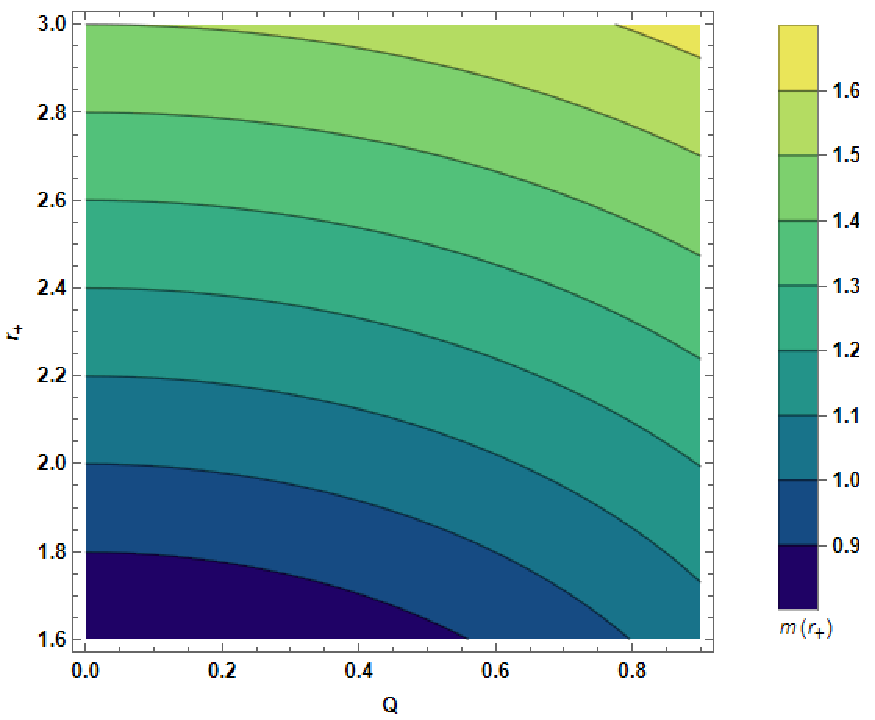}
\end{tabular}
\end{center}
\caption{Contour plot of mass $m(r_+)$ against the charge $Q$ and outer event horizon $r_+$ corresponding to $\kappa\lambda = -3/2$, $\mathcal{N}_q = 0.1$ (Left) and $\kappa\lambda = 3/2$, $\mathcal{N}_q = -0.1$ (Right).}\label{fig2}
\end{figure}

Also, the equality behaviours of $H^0_{~~0} = H^1_{~~1}$ and $H^2_{~~2} = H^3_{~~3}$ restrict $T^0_{~~0} = T^1_{~~1}$ and $T^2_{~~2} = T^3_{~~3}$.  Therefore, a general total energy-momentum tensor $T^\mu_{~~\nu}$ possessing these symmetry properties can be constructed as 
\begin{equation}
    T^\mu_{~~\nu} = E^\mu_{~~\nu}+\mathcal{T}^\mu_{~~\nu},\label{TT}
\end{equation}
where $E^\mu_{~~\nu}$ is the trace-free Maxwell tensor and $\mathcal{T}^\mu_{~~\nu}$ is the energy-momentum tensor of the surrounding field. The trace-free Maxwell tensor $E^\mu_{~~\nu}$ can be written as
\begin{eqnarray}
    E_{\mu\nu} =\frac{2}{\kappa}\left(F_{\mu\alpha}F_\nu^{~\alpha}-\frac{1}{4}g_{\mu\nu}F^{\alpha\beta}F_{\alpha\beta}\right),\label{E}
\end{eqnarray}

where $F_{\alpha\beta}$ is the antisymmetric Faraday tensor that satisfies the vacuum Maxwell equations given by
\begin{eqnarray}
 F^{\mu\nu}_{~~~;\nu} &=& 0,
 \\
 \partial_{[\sigma F_{\mu\nu}]} &=& 0.\label{F}
\end{eqnarray}

The non-vanishing components of the Faraday tensor $F_{\alpha \beta}$ for the spherical symmetry spacetime metric (\ref{ds}) satisfy $F^{01}$ = -$F^{10}$. Therefore, Eq. (\ref{F}) yields the following result
\begin{eqnarray}
    F^{01} = \frac{Q}{r^2},\label{F1}
\end{eqnarray}
where $Q$ is an integration constant that plays the role of an electrostatic charge. Thus, from Eqns. (\ref{ds}), (\ref{E}) and (\ref{F1}), one can obtain the non-zero components of the Maxwell tensor $E^\mu_{~~\nu}$ as
\begin{eqnarray}
   E^\mu_{~~\nu} = \frac{Q^2}{\kappa r^4} \text{diag}\left(-1,~ -1,~ 1,~ 1\right).
\end{eqnarray}

\begin{figure}[!htbp]
\begin{center}
\begin{tabular}{rl}
\includegraphics[width=8.3cm]{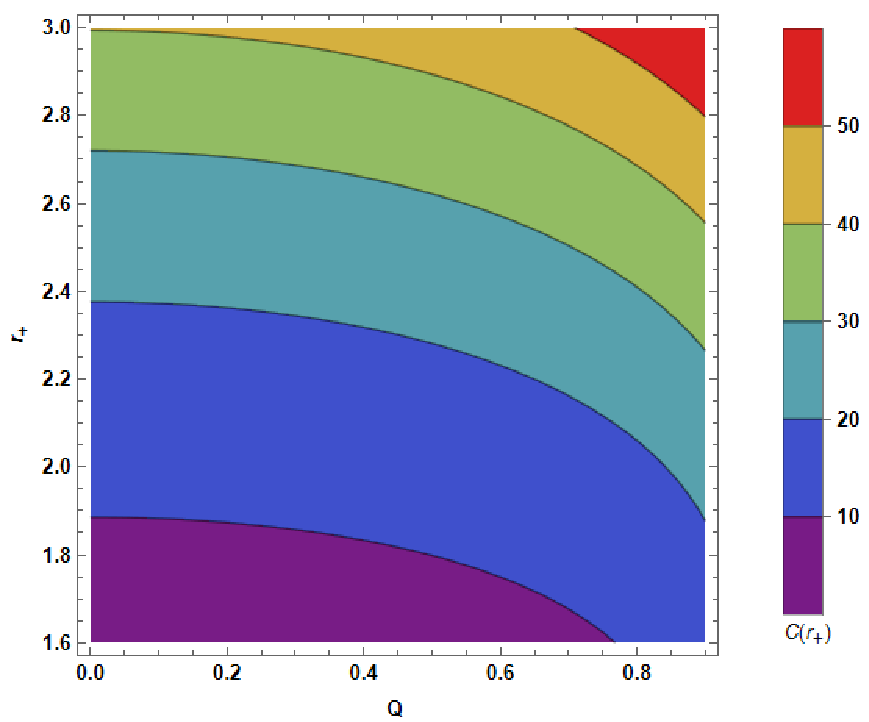}
\end{tabular}
\begin{tabular}{rl}
\includegraphics[width=8.1cm]{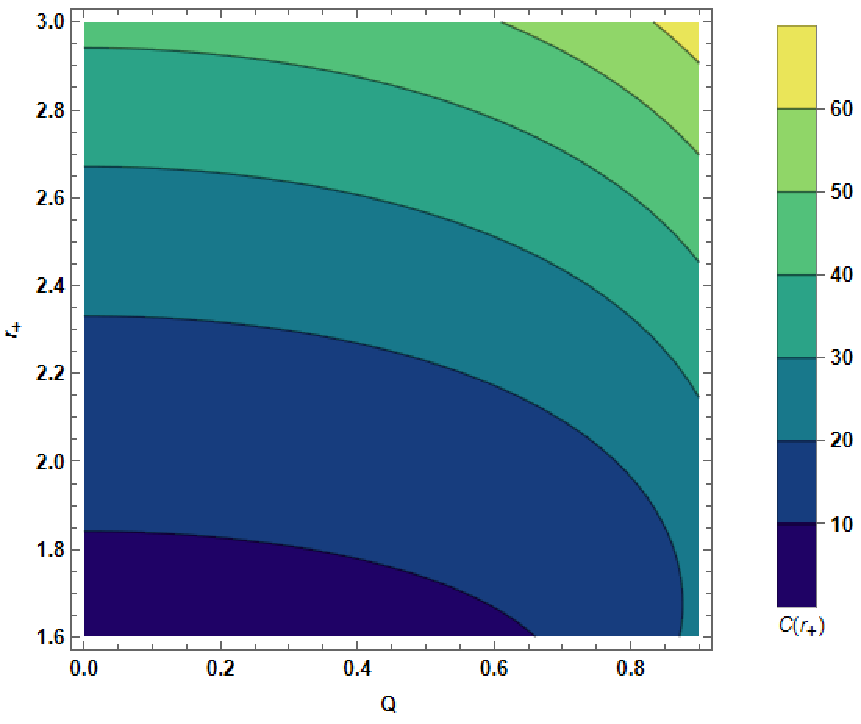}
\end{tabular}
\end{center}
\caption{Contour plot of heat capacity $C(r_+)$ against the charge $Q$ and outer event horizon $r_+$ corresponding to $\kappa\lambda = -3/2$, $\mathcal{N}_q = 0.1$, $m$ = 1 (Left) and $\kappa\lambda = 3/2$, $\mathcal{N}_q = -0.1$, $m$ = 1 (Right).}\label{fig3}
\end{figure}

The energy-momentum tensor of the surrounding field $\mathcal{T}^\mu_{~~\nu}$ can be expressed as \cite{vv03}
\begin{eqnarray}
    \mathcal{T}^0_{~~0} &=& -\rho_s(r),
    \\
    \mathcal{T}^i_{~~j} &=& -\alpha\rho_s(r)\left[-\left(1+3\beta\right)\frac{r_ir^j}{r_nr^n}+\beta \delta^i_j\right],
\end{eqnarray}
where $\rho_s(r)$ represents the energy density of the surrounding field of the black hole and the parameters $\alpha$ and $\beta$ are associated with the internal structure of the surrounding field. In this regard,  the isotropic scenario gives the following result \cite{vv03}
\begin{eqnarray}
    \langle T^i_{~~j} \rangle = \frac{\alpha}{3} \rho_s \delta^i_{~~j} = p_s \delta^i_{~~j}, ~~~\text{as} ~~~\langle r^ir_j\rangle = \frac{1}{3}\delta^i_{~~j} r_n r^n. 
\end{eqnarray}
where $p_s$ is the pressure of the surrounding field. Thus, the surrounding field obeys a barotropic equation of state, defined as 
\begin{equation}
    p_s = \omega_s\rho_s,~~~\omega_s = \frac{1}{3}\alpha,
\end{equation}
where $\omega_s$ is the equation of state parameter. Indeed, the field equations (\ref{h1})-(\ref{h4}) along with the total energy-momentum tensor (\ref{TT}) can describe the principle of additivity and linearity condition \cite{vv03} that helps to determine the free parameter $\beta$ of the energy-momentum tensor for the surrounding field as follows
\begin{eqnarray}
    \beta = -\frac{1+3\omega_s}{6\omega_s}.
\end{eqnarray}

Thus, the non-zero components of the energy-momentum tensor of the surrounding field $\mathcal{T}^\mu_\nu$ can be obtained as
\begin{eqnarray}
    \mathcal{T}^0_{~~0} &=& \mathcal{T}^1_{~~1} = -\rho_s(r),
    \\
    \mathcal{T}^2_{~~2} &=& \mathcal{T}^3_{~~3} = \frac{1}{2}\left(1+3\omega_s\right)\rho_s.
\end{eqnarray}

Now, the $H^0_{~~0}$ = $T^0_{~~0}$ and $H^1_{~~1}$ = $T^1_{~~1}$ components of the Rastall field equations processes the following result 
\begin{eqnarray}
\frac{1}{r^2}\left(r\mathcal{F}'-1+\mathcal{F}\right) -\frac{\kappa\lambda}{r^2}\left(r^2\mathcal{F}''+ 4r\mathcal{F}'-2+2\mathcal{F}\right) = -\kappa\rho_s-\frac{Q^2}{r^4},\label{D1}
\end{eqnarray}
and the $H^2_{~~2}$ = $T^2_{~~2}$ and $H^3_{~~3}$ = $T^3_{~~2}$ components give the following result
\begin{eqnarray}
    \frac{1}{r^2}\left(r\mathcal{F}'+\frac{r^2}{2}\mathcal{F}''\right)-\frac{\kappa\lambda}{r^2}\left(r^2\mathcal{F}"+ 4r\mathcal{F}'-2+2\mathcal{F}\right) =\frac{1}{2}\left(1+3\omega_s\right)\kappa\rho_s+\frac{Q^2}{r^4}.\label{D2}
\end{eqnarray}

\begin{figure}[!htbp]
\begin{center}
\begin{tabular}{rl}
\includegraphics[width=8cm]{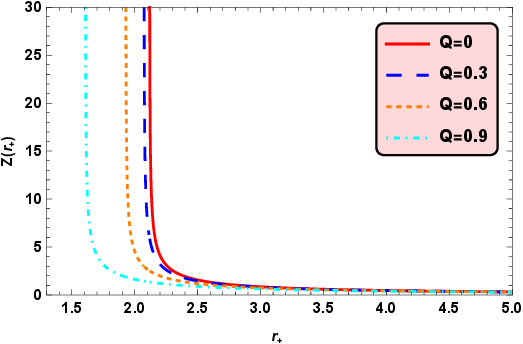}
\end{tabular}
\begin{tabular}{rl}
\includegraphics[width=8cm]{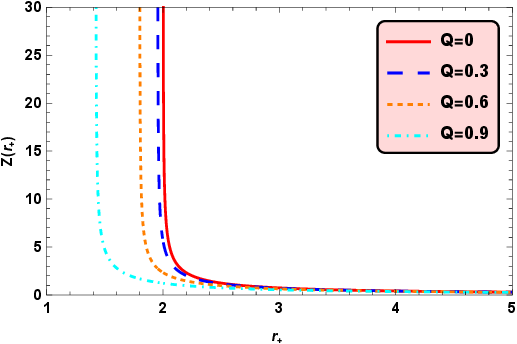}
\end{tabular}
\end{center}
\caption{Profiles of the gravitational redshift $Z(r_+)$ against outer event horizon $r_+$ corresponding to $\kappa\lambda = -3/2$, $\mathcal{N}_q = 0.1$,  $m$ = 1 (Left) and $\kappa\lambda = 3/2$, $\mathcal{N}_q = -0.1$,  $m$ = 1 (Right).}\label{fig4}
\end{figure}

Solving the above Eqs. (\ref{D1}) and (\ref{D2}), the general solution for the metric function $\mathcal{F}(r)$ and energy density $\rho_s$ can be obtained as 
\begin{eqnarray}
    \mathcal{F}(r) = 1-\frac{2 m }{r}+\frac{Q^2}{r^2}-\frac{\mathcal{N}_s}{r^{\frac{1+3\omega_s-6\kappa\lambda(1+\omega_s)}{1-3\kappa\lambda(1+\omega_s)}}},
\end{eqnarray}
and 
\begin{eqnarray}
    \rho_s = -\frac{3\mathcal{W}_s\mathcal{N}_s}{\kappa r^{\frac{3(1+\omega_s)-12\kappa\lambda(1+\omega_s)}{1-3\kappa\lambda(1+\omega_s)}} },
\end{eqnarray}

where $m$ and $\mathcal{N}_s$ are two integration constants that denote the black hole mass and the parameter characterizing the surrounding field structure, respectively. Here,
\begin{eqnarray}
    \mathcal{W}_s = -\frac{(1-4\kappa \lambda)[\kappa\lambda(1+\omega_s)-\omega_s]}{[1-3\kappa\lambda (1+\omega_s)]^2},\label{W}
\end{eqnarray}
represents a geometric constant that depends on $\kappa$, $\lambda$ and $\omega_s$. It is important to note that the satisfaction of weak energy condition demands the positive energy density of the surrounding field i.e. $\rho_s \geq 0$, which ensures $\mathcal{W}_s\mathcal{N}_s \leq 0$. Therefore, if the surrounding field has $\mathcal{W}_s > 0$, then $\mathcal{N}_s$ must be less than zero and conversely for $\mathcal{W}_s < 0$, we need $\mathcal{N}_s > 0$. Indeed, the result (\ref{W}) for $\mathcal{W}_s$ clearly indicates that the sing of the surrounding field parameter $\mathcal{N}_s$ depends on $\kappa$, $\lambda$ and $\omega_s$. Finally, the surrounded charged black hole solution takes the following form:
\begin{eqnarray}
    ds^2 = -\left(1-\frac{2 m }{r}+\frac{Q^2}{r^2}-\frac{\mathcal{N}_s}{r^{\frac{1+3\omega_s-6\kappa\lambda(1+\omega_s)}{1-3\kappa\lambda(1+\omega_s)}}}\right) dt^2+\frac{dr^2}{1-\frac{2 m }{r}+\frac{Q^2}{r^2}-\frac{\mathcal{N}_s}{r^{\frac{1+3\omega_s-6\kappa\lambda(1+\omega_s)}{1-3\kappa\lambda(1+\omega_s)}}}}+r^2(d\theta^2+\sin\theta d\phi^2).\label{DS1}
\end{eqnarray}
It is noted that for $\lambda = 0$ and $\kappa = 8\pi G_s$, one can get the Reissner–Nordström black hole surrounded by a surrounding field in general relativity and this is known as the  Kiselev black hole solution \cite{vv03}.

Now, as we are looking for a charged black hole solution surrounded by the quintessence field in Rastall’s theory of gravity, we set $\omega_s$ = $\omega_q$ = -2/3 \cite{vv03, AV05}. Thus, for a charged black hole solution surrounded by the quintessence field in Rastall gravity, the metric (\ref{DS1}) reads as 
\begin{equation}
    ds^2 = -\mathcal{F}(r)dt^2+\frac{1}{\mathcal{F}(r)}dr^2+r^2(d\theta^2+\sin\theta d\phi^2).\label{bh}
\end{equation}
where,
\begin{eqnarray}
    \mathcal{F}(r) = 1-\frac{2 m }{r}+\frac{Q^2}{r^2}-\frac{\mathcal{N}_q}{r^{\frac{-1-2\kappa\lambda}{1-\kappa\lambda}}}.
\end{eqnarray}

It can be observed that the above metric differs from the metric of a charged black hole surrounded by a quintessence field in general relativity \cite{vv03}. Indeed, for $\kappa\lambda \neq 0$, the geometric parameters $\kappa$ and $\lambda$ in Rastall gravity can significantly influence the solutions, resulting in differences compared to those in general relativity. Also, for $Q = 0$, i.e. $E^\mu_{~~\nu} = 0$ in the total energy-momentum tensor (\ref{TT}), one can obtain the uncharged Kiselev-like black hole solutions within the quintessence field. Moreover, it is important to note that the spacetime of the black hole (\ref{bh}) becomes asymptotically flat when $\kappa\lambda < -1/2$ or $\kappa\lambda >1$.

\begin{figure}[!htbp]
\begin{center}
\begin{tabular}{rl}
\includegraphics[width=8cm]{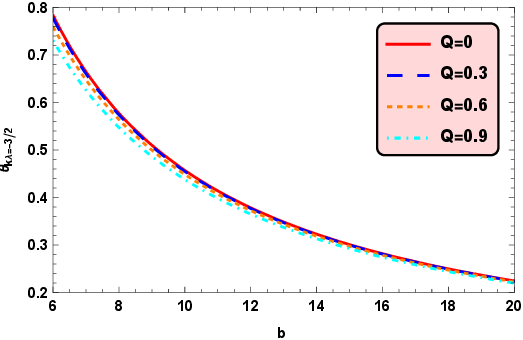}
\end{tabular}
\begin{tabular}{rl}
\includegraphics[width=8cm]{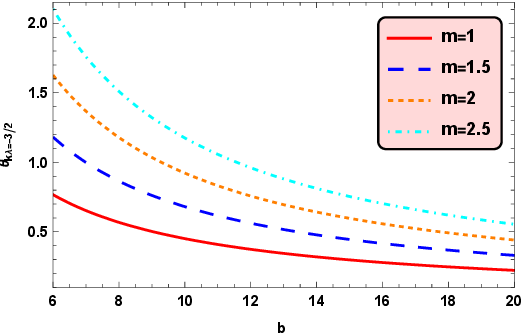}
\end{tabular}
\end{center}
\caption{Profiles of the deflection angle $\theta_{\kappa\lambda=-3/2}$ against the impact parameter $b$ corresponding to $\mathcal{N}_q = 0.1$ along with $m = 1$ (Left) and  $\mathcal{N}_q = 0.1$ along with $Q = 0.5$ (Right).}\label{fig5}
\end{figure}

In this case, the geometric parameter $\mathcal{W}_s$ = $\mathcal{W}_q$ (\ref{W}) takes the following form
\begin{eqnarray}
    \mathcal{W}_q = -\frac{(1-4\kappa \lambda)(2+\kappa\lambda)}{3(1-\kappa\lambda)^2},\label{W1}
\end{eqnarray}
Therefore, in this case, $\mathcal{W}_q < 0$ for $-2 <\kappa\lambda <1/4$ and $\mathcal{W}_q > 0$ for $\kappa\lambda >1/4$.  Consequently, to satisfy the weak energy condition, $\mathcal{N}_q > 0$ for $-2 <\kappa\lambda <1/4$ and $\mathcal{N}_q < 0$ for $\kappa\lambda >1/4$. In this study, we consider two cases: $\kappa\lambda = -3/2$ \cite{yh17} with $\mathcal{N}_q = 0.1$ and $\kappa\lambda = 3/2$ \cite{yh17} with $\mathcal{N}_q = -0.1$, chosen to be consistent with both the asymptotic nature of the spacetime and the weak energy condition. In order to examine the event horizons of the black hole (\ref{bh}) we demonstrate $\mathcal{F}(r)$ graphically in Fig. \ref{fig1} for both the cases: $\kappa\lambda = -3/2$, $\mathcal{N}_q = 0.1$  and $\kappa\lambda = 3/2$, $\mathcal{N}_q = -0.1$. For the case: $\kappa\lambda = -3/2$, $\mathcal{N}_q = 0.1$, the present black hole has only one event horizon corresponding to $Q = 0$, and hence, the uncharged Kiselev-like black hole solution within the quintessence field has one event horizon like the Schwarzchild black hole but it larger than the event horizon of  Schwarzchild black hole. However, for $Q \neq 0$, the black hole solutions have both the inner and outer event horizons. In the charge black holes scenario, the inner horizons increase for increasing values of charge $Q$ and decrease for increasing values of mass $m$, whereas the outer horizons decrease for increasing values of charge $Q$ and increase for increasing values of mass $m$ (See Fig. \ref{fig1} (Left)). For the case: $\kappa\lambda = 3/2$, $\mathcal{N}_q = -0.1$, both inner and outer event horizons have existed for uncharged as well as charged black holes and they maintain the same behaviours against the charge $Q$ and mass $m$ as of the previous case, clear from Fig. \ref{fig1} (Right). In addition, the numerical values of inner and outer event horizons $r_{\mp}$ are provided in Tables- \ref{tab1} and \ref{tab2}.

Now, we explore some thermodynamic features of the present black hole solutions (\ref{bh}). In this regard, the mass of the black hole is obtained from the condition $g_{tt}(r_\pm) = 0$ as
\begin{eqnarray}
    m(r_+) = \frac{1}{2 r_+}\left[r_+^2+Q^2-\mathcal{N}_q r_+^{\frac{3}{1-k \lambda }}\right].
\end{eqnarray}

The profiles of mass $m(r_+)$ are displayed in Fig. \ref{fig2} for the cases   $\kappa\lambda$ = -3/2, $\mathcal{N}_q$ = 0.1 (Left) and $\kappa\lambda$ = 3/2, $\mathcal{N}_q$ = -0.1 (Right), both the figures indicate that the mass $m(r_+)$ increases for increasing values of $r_+$.

 The surface gravity $\mathcal{S}_g$, Hawking temperature $\mathcal{H}_t$, and the heat capacity $C$ of the  black hole is obtained as \cite{sc98}
 \begin{eqnarray}
    \mathcal{S}_g(r_+) &=& \frac{1}{2}\left[\frac{d g_{tt}}{dr}\right]_{r_+} = \frac{1}{2  r_+^3 (k \lambda -1)}\left[2 m r_+ (k \lambda -1)+2Q^2 (1-k \lambda )+\mathcal{N}_q (2 k \lambda +1) r_+^{\frac{3}{1-k \lambda }}\right],
\\
    \mathcal{H}_t(r_+) &=& \frac{1}{4\pi}\left[\frac{d g_{tt}}{dr}\frac{1}{\sqrt{-g_{tt}g_{rr}}}\right]_{r_+} = \frac{1}{4 \pi  r_+^3 (k \lambda -1)}\left[2 m r_+ (k \lambda -1)+2Q^2 (1-k \lambda )+\mathcal{N}_q (2 k \lambda +1) r_+^{\frac{3}{1-k \lambda }}\right],
\\
    C(r_+) &=& \frac{dm(r_+)}{dT(r_+)} = \frac{2 \pi  r^2 (k \lambda -1) \left[\mathcal{N}_q (k \lambda +2) r^{\frac{3}{1-k \lambda }}-k \lambda  Q^2+r^2 (k \lambda -1)+q\right]}{4 m r (k \lambda -1)^2+3 k \lambda  \mathcal{N}_q (2 k \lambda +1) r^{\frac{3}{1-k \lambda }}-6 Q^2 (k \lambda -1)^2}.
\end{eqnarray}

For both the considered values of the Rastal gravity parameter, we demonstrate the profiles of heat capacity $C(r_+)$ in Fig. \ref{fig3}, which ensures that $C(r_+)$ is positive. Consequently, the present black hole solutions are always stable.

 Also, the gravitational redshift of the present black hole can be written as 
\begin{eqnarray}
    Z(r) = [\mathcal{F}(r)]^{-1/2}-1 = \left[1-\frac{2 m }{r}+\frac{Q^2}{r^2}-\frac{\mathcal{N}_q}{r^{\frac{-1-2\kappa\lambda}{1-\kappa\lambda}}}\right]^{-1/2}-1.
\end{eqnarray}

The behaviour of $Z(r)$ is illustrated in Fig. \ref{fig4}, showing that it increases as it approaches the black hole's singularity as desired.  We will now estimate the deflection angle of light in the weak-field approximation for the considered black hole (\ref{bh}) using the Gauss-Bonnet method.

\begin{figure}[!htbp]
\begin{center}
\begin{tabular}{rl}
\includegraphics[width=8cm]{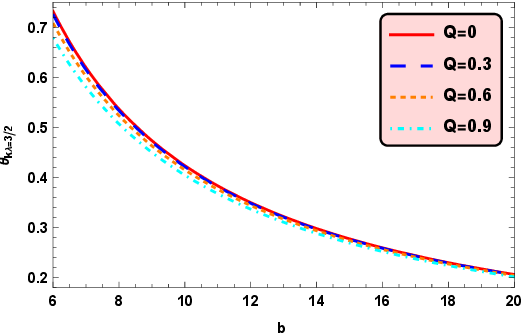}
\end{tabular}
\begin{tabular}{rl}
\includegraphics[width=8cm]{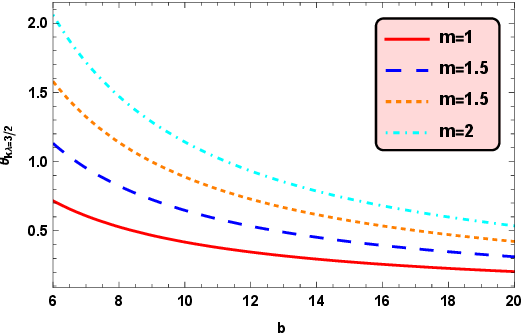}
\end{tabular}
\end{center}
\caption{Profiles of the deflection angle $\theta_{\kappa\lambda=3/2}$ against the impact parameter $b$ corresponding to  $\mathcal{N}_q = -0.1$ along with $m = 1$ (Left) and  $\mathcal{N}_q = -0.1$ along with $Q = 0.5$ (Right).}\label{fig6}
\end{figure}

\section{Formulation of Defection angle: Gauss-Bonnet method}\label{sec3}
In this section, we will present the mathematical formula for the weak deflection angle of light in the asymptotically flat spacetime of a massive celestial object. In this context, Gibbons and Werner \cite{gibbons1} proposed a highly effective method for estimating the deflection angle of light in the weak-field approximation for a massive celestial object with asymptotic spacetime behaviour by describing the relationship between the intrinsic geometry of the spacetime and the topology of its non-singular regions.

{\bf{Theorem:}} Let $\mathcal{S}_{R}$  be a non-singular region surrounded by the beam of light with boundary $\partial \mathcal{S}_{R}$ = $\gamma _{g^{op}}\cup \gamma_R$ and let $\mathcal{G}$ be the Gaussian optical curvature and $\mathcal{C}$ be the geodesic curvature. Then the Gauss-Bonnet theorem is stated as \cite{gibbons1}
\begin{equation}\label{GBT}
\iint\limits_{\mathcal{S}_{R}}\mathcal{G}\,\mathrm{d}\mathcal{S}+\oint\limits_{\partial \mathcal{%
S}_{R}}\mathcal{C} \,\mathrm{d}t+\sum_{i}\theta _{i}=2\pi \chi (\mathcal{S}_{R}),
\end{equation}

where $\mathrm{d}\mathcal{S}$ is the surface element, $\theta_{i}$ is the exterior angle at the $i^{th}$ vertex, and $\chi$ is the Euler characteristic element with $\chi (\mathcal{S}_{R}) = 1$.

Now, the geodesic curvature $\mathcal{C}$ is defined as
\begin{equation}
\mathcal{C} =g^{\text{op}}\,\left(\nabla _{\dot{%
\gamma}}\dot{\gamma},\ddot{\gamma}\right)~~\text{with} ~~ g^{op}(\dot{\gamma},\dot{%
\gamma}) = 1,
\end{equation}
where $\ddot{\gamma}$ represents the unit acceleration vector. It is important to note that the sum of jump angles $\theta_v$ and $\theta_s$ to the viewer and source, respectively become $\pi$ for $R\rightarrow \infty$ i.e. $\theta _v$ + $ \theta _s\rightarrow \pi$  when $R\rightarrow \infty$ \cite{gibbons1}. Additionally, for a geodesic $\gamma _{g^{op}}$, we have $\mathcal{C}(\gamma _{g^{op}})=0$. Thus, the geodesic curvature  can be expressed as
\begin{equation}
\mathcal{C} (\gamma_{R})=|\nabla _{\dot{\gamma}_{R}}\dot{\gamma}_{R}|,
\end{equation}

The radial part of the geodesic curvature can be written as
\begin{equation}
\left( \nabla _{\dot{\gamma}_{R}}\dot{\gamma}_{R}\right) ^{r}=\dot{\gamma}_{R}^{\varphi
}\,\left( \partial _{\varphi }\dot{\gamma}_{R}^{r}\right) +\tilde{\Gamma} _{\varphi
\varphi }^{r}\left( \dot{\gamma}_{R}^{\varphi }\right) ^{2}, \label{12}
\end{equation}

with $\gamma_{R}:=r(\varphi)=R=\text{constant}$ for large $R$, and,  $\tilde{\Gamma}_{\varphi\varphi }^{r}$ stands for the Christoffel symbol of the optical metric. Now, one can see that the first term of the preceding equation becomes zero as it does not involve the topological effect whereas the second term can be determined with the aid of the unit speed condition $\tilde{g}_{\varphi \varphi}\dot{\gamma}_{R}^{\varphi } \dot{\gamma}_{R}^{\varphi }=1 $. Consequently, the geodesic curvature reads as
\begin{eqnarray}\notag\label{gcurvature}
\lim_{R\rightarrow \infty }\mathcal{C} (\gamma_{R}) &=&\lim_{R\rightarrow \infty
}\left\vert \nabla _{\dot{\gamma}_{R}}\dot{\gamma}_{R}\right\vert  \notag 
\rightarrow \frac{1}{R}.
\end{eqnarray}%

\begin{figure}[!htbp]
\begin{center}
\begin{tabular}{rl}
\includegraphics[width=8cm]{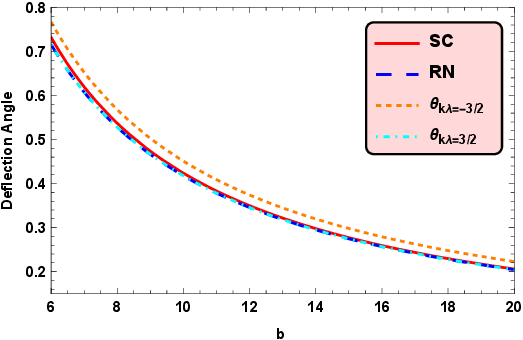}
\end{tabular}
\end{center}
\caption{Profiles of the Schwarzchild black hole's deflection angle (SC), Reissner–Nordstrom black hole's deflection angle (RN) and the deflection angles $\theta_{\kappa\lambda=3/2}$, and $\theta_{\kappa\lambda=-3/2}$ against the impact parameter $b$ corresponding to $m = 1$, $Q = 0.5$.}\label{fig7}
\end{figure}

Therefore, for a sufficiently large $R$, we obtained the result  $\kappa (\gamma_{R}) dt$ = $d\phi$. Finally, the Gauss-Bonnet theorem (\ref{GBT}) can be expressed as
\begin{equation}
\iint\limits_{\mathcal{S}_{R}}\mathcal{G}\,\mathrm{d}\mathcal{S}+\oint\limits_{\gamma_{R}}\mathcal{C} \,%
\mathrm{d}t\overset{{R\rightarrow \infty }}{=}\iint\limits_{\mathcal{S}%
_{\infty }}\mathcal{G}\,\mathrm{d}\mathcal{S}+\int\limits_{0}^{\pi + \theta_A}\mathrm{d}\varphi
=\pi.\label{A}
\end{equation}

Now, in the scenario of the light beam that follows a straight line approximation  $r_\gamma = b/ \sin\phi$, where $b$ is the impact parameter the above deflection (\ref{A}) takes the following form
\begin{eqnarray}\label{GBT2}
\theta_A &=&-\int\limits_{0}^{\pi}\int\limits_{b/ \sin\phi}^{\infty} \mathcal{G} \mathrm{d}\mathcal{S}.\label{DA}
\end{eqnarray}
It is important to note here that one can approximate the impact parameter $b$ up to the first order by the closest distance to the black hole.

\section{Deflection angle of the black hole} \label{sec4}
In order to estimate the deflection angle, we take into account the optical metric in the gravitational field of the black hole (\ref{bh}) that can be described by null geodesics equations $ds^2$ = 0 at the equatorial plane $\theta = \pi/2$. Therefore, the optical metric of the black hole (\ref{bh}) reads as
\begin{equation}
dt^2=\left[1-\frac{2 m }{r}+\frac{Q^2}{r^2}-\frac{\mathcal{N}_q}{r^{\frac{-1-2\kappa\lambda}{1-\kappa\lambda}}}\right]^{-2}dr^2+r^2\left[1-\frac{2 m }{r}+\frac{Q^2}{r^2}-\frac{\mathcal{N}_q}{r^{\frac{-1-2\kappa\lambda}{1-\kappa\lambda}}}\right]^{-1} d\phi^2.\label{bh1}
\end{equation}

Additionally, the optical metric can be expressed in terms of the  Regge-Wheeler tortoise static radial coordinate $r^\star$ in the following form
\begin{equation}
\mathrm{d}t^2 \equiv g_{ab}^{op} \mathrm{d}x^a \mathrm{d}x^b={\mathrm{d}r^{\star}}^{2}+{f(r^{\star})}^2 \mathrm{d}\varphi^2.\label{rw}
\end{equation}

Therefore, comparing  Eqs. (\ref{bh1}) and (\ref{rw}) one can obtain the following results
\begin{eqnarray}
\mathrm{d}r^\star &=&\left[1-\frac{2 m }{r}+\frac{Q^2}{r^2}-\frac{\mathcal{N}_q}{r^{\frac{-1-2\kappa\lambda}{1-\kappa\lambda}}}\right]^{-1}dr,
\end{eqnarray}
\begin{eqnarray}
f(r^\star)&=& r \left[1-\frac{2 m }{r}+\frac{Q^2}{r^2}-\frac{\mathcal{N}_q}{r^{\frac{-1-2\kappa\lambda}{1-\kappa\lambda}}}\right]^{-\frac{1}{2}}.
\end{eqnarray}

\begin{figure}[!htbp]
\begin{center}
\begin{tabular}{rl}
\includegraphics[width=8cm]{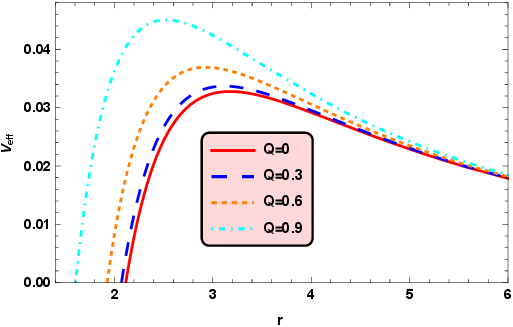}
\end{tabular}
\begin{tabular}{rl}
\includegraphics[width=8cm]{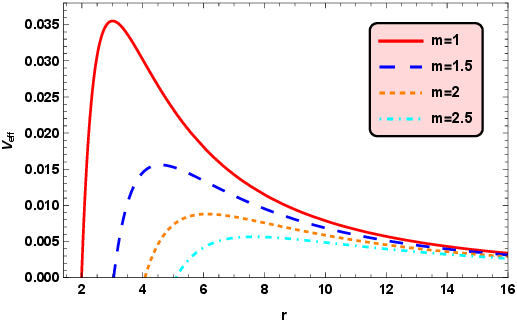}
\end{tabular}
\end{center}
\caption{Profiles of the effective potential against the radial coordinate $r$ corresponding to $\kappa\lambda = -3/2$ along with $\mathcal{N}_q = 0.1$, $m = 1$ (Left) and $\kappa\lambda = -3/2$ along with $\mathcal{N}_q = 0.1$, $Q = 0.5$ (Right).}\label{fig8}
\end{figure}

Also, the  Gaussian curvature $\mathcal{G}$ of the optical surface can be expressed in terms of the  Regge-Wheeler tortoise static radial coordinate $r^\star$ as \cite{gibbons1}
\begin{eqnarray}
\mathcal{G}&=&-\frac{1}{f(r^{\star })}\frac{\mathrm{d}^{2}f(r^{\star })}{\mathrm{d}{%
r^{\star }}^{2}}= -\frac{1}{f(r^{\star })}\left[ \frac{\mathrm{d}r}{\mathrm{d}r^{\star }}%
\frac{\mathrm{d}}{\mathrm{d}r}\left( \frac{\mathrm{d}r}{\mathrm{d}r^{\star }}%
\right) \frac{\mathrm{d}f}{\mathrm{d}r}+\left( \frac{\mathrm{d}r}{\mathrm{d}%
r^{\star }}\right) ^{2}\frac{\mathrm{d}^{2}f}{\mathrm{d}r^{2}}\right].\label{g}
\end{eqnarray}

By applying the above formula (\ref{g}),  the optical Gaussian curvature for the optical metric (\ref{bh1}) is obtained as
\begin{eqnarray}
\mathcal{G} &=& \frac{1}{4 r^6 (k \lambda -1)^2}\big[4 (k \lambda -1)^2 \left\{3 Q^2 r (r-2 m)+m r^2 (3 m-2 r)+2 Q^4\right\}+\mathcal{N}_q^2 (2 k \lambda +1) (4 k \lambda -1) r^{-\frac{6}{k \lambda -1}}\nonumber
\\
&&+2 \mathcal{N}_q r^{\frac{3}{1-k \lambda }} \{Q^2 (k \lambda  (7-8 k \lambda )-8)-3 k \lambda  r (2 k \lambda  (r-2 m)+r)+6 m r\}\big].\label{cur}
\end{eqnarray}

Also, the surface element $d\mathcal{S}$ for the optical metric (\ref{bh1}) can be expressed as
\begin{eqnarray}
    d\mathcal{S} = \sqrt{g^{opt}}dr d\phi = r \left[1-\frac{2 m }{r}+\frac{Q^2}{r^2}-\frac{\mathcal{N}_q}{r^{\frac{-1-2\kappa\lambda}{1-\kappa\lambda}}}\right]^{-\frac{3}{2}} dr d\phi.
\end{eqnarray}

Now, we will proceed to determine the deflection angle by taking two special values of Rastall gravity parameter $\kappa\lambda$.

\subsection{Case-I: $\kappa\lambda$ = -3/2}
In this section, we will determine the deflection angle for the choice of Rastall gravity parameter $\kappa\lambda$ = -3/2. For this value of $\kappa\lambda$, we consider the surface element in the following approximation form
\begin{eqnarray}
    d\mathcal{S}_{\kappa\lambda= -\frac{3}{2}} = r \left[1-\frac{2 m }{r}+\frac{Q^2}{r^2}-\frac{\mathcal{N}_q}{r^{\frac{4}{5}}}\right]^{-\frac{3}{2}} dr d\phi\simeq r\left[1+\frac{3 \mathcal{N}_q}{2 r^{4/5}}+\frac{3 m}{r}+\frac{15 \mathcal{N}_q^2}{6 r^{8/5}}\right] dr d\phi.\label{ds1}
\end{eqnarray}

In this case, the  optical Gaussian curvature (\ref{cur}) reads as
\begin{eqnarray}
\mathcal{G}_{\kappa\lambda = -\frac{3}{2}} &=& -\frac{18 \mathcal{N}_q}{25 r^{14/5}}-\frac{2 m}{r^3}+\frac{14 \mathcal{N}_q^2}{25 r^{18/5}}+\frac{66 m \mathcal{N}_q}{25 r^{19/5}}+\frac{3(m^2+ Q^2)}{r^4}-\frac{73 \mathcal{N}_q Q^2}{25 r^{24/5}}-\frac{6 m Q^2}{r^5}+\frac{2 Q^4}{r^6}.\label{cur1}
\end{eqnarray}

\begin{figure}[!htbp]
\begin{center}
\begin{tabular}{rl}
\includegraphics[width=8cm]{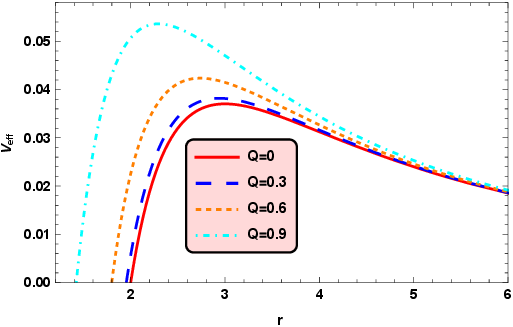}
\end{tabular}
\begin{tabular}{rl}
\includegraphics[width=8cm]{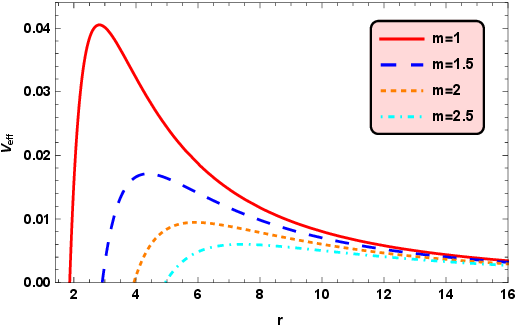}
\end{tabular}
\end{center}
\caption{Profiles of the effective potential against the radial coordinate $r$ corresponding to  $\kappa\lambda = 3/2$ along with $\mathcal{N}_q = -0.1$, $m = 1$ (Left) and $\kappa\lambda = 3/2$ along with $\mathcal{N}_q = -0.1$, $Q = 0.5$ (Right).}\label{fig9}
\end{figure}

Therefore, on using Eqns. (\ref{ds1}) and (\ref{cur1}) in Eq. (\ref{DA}) we obtain the deflection angle in the following form
\begin{eqnarray}
    \Theta_{\kappa\lambda = -\frac{3}{2}} &\simeq& \frac{4 m}{b}+ \frac{3 \pi  \left(m^2-Q^2\right)}{4 b^2}+\Bigg[\frac{16576 \pi ^{3/2} \sqrt{\frac{2}{\sqrt{5}+5}} m Q^2}{5415 b^{19/5} \Gamma \left(\frac{3}{5}\right) \Gamma \left(\frac{9}{10}\right)}+\frac{56 \pi ^{3/2} \sqrt{\frac{2}{\sqrt{5}+5}} m}{45 b^{9/5} \Gamma \left(\frac{3}{5}\right) \Gamma \left(\frac{9}{10}\right)}-\frac{9 \pi ^{3/2} \left(621 m^2+79 Q^2\right)}{490 \left(\sqrt{5}-1\right) b^{14/5} \Gamma \left(\frac{1}{10}\right) \Gamma \left(\frac{7}{5}\right)}\nonumber
    \\
    &&-\frac{\left(\sqrt{5}+1\right) \pi ^{3/2}}{b^{4/5} \Gamma \left(-\frac{9}{10}\right) \Gamma \left(\frac{7}{5}\right)}-\frac{285 \pi ^{3/2} Q^4}{224 \left(\sqrt{5}-1\right) b^{24/5} \Gamma \left(\frac{1}{10}\right) \Gamma \left(\frac{7}{5}\right)}\Bigg]\mathcal{N}_q+ \Bigg[\frac{2160 \pi ^{3/2} \sqrt{10 \left(\sqrt{5}+5\right)} m Q^2}{6877 b^{23/5} \Gamma \left(\frac{1}{5}\right) \Gamma \left(\frac{13}{10}\right)}\nonumber
    \\
    &&-\frac{13 \pi ^{3/2} \left(125 m^2+52 Q^2\right)}{720 \left(\sqrt{5}+1\right) b^{18/5} \Gamma \left(\frac{7}{10}\right) \Gamma \left(\frac{4}{5}\right)}-\frac{128 \sqrt{2+\frac{2}{\sqrt{5}}} \pi ^{3/2} m}{845 b^{13/5} \Gamma \left(\frac{1}{5}\right) \Gamma \left(\frac{13}{10}\right)}-\frac{7475 \pi ^{3/2} Q^4}{9408 \left(\sqrt{5}+1\right) b^{28/5} \Gamma \left(\frac{7}{10}\right) \Gamma \left(\frac{4}{5}\right)}\nonumber
    \\
    &&+\frac{39 \pi ^{3/2}}{80 \left(\sqrt{5}+1\right) b^{8/5} \Gamma \left(\frac{7}{10}\right) \Gamma \left(\frac{4}{5}\right)}\Bigg]\mathcal{N}_q^2.
\end{eqnarray}

From the above result, one can see that the deflection angle is influenced by the surrounding field structure parameter $\mathcal{N}_q$. For $\mathcal{N}_q$ = 0, the deflection reduces to $4m/b$+ $3\pi\left(m^2-Q^2\right)/4 b^2$, which is the deflection angle of the Reissner–Nordstrom black hole \cite{yk22, gy22} in second order approximation. Moreover, the obtained deflection angle reduces to $4m/b$ + $3\pi m^2/4 b^2$ for $\mathcal{N}_q$ = 0 and $Q$ = 0  that matched with the deflection angle of the Schwarzchild black hole \cite{yk22, jb03}.

\subsection{Case-II: $\kappa\lambda$ = 3/2}
Here, we estimate the deflection angle for the Rastall gravity parameter $\kappa\lambda$ = 3/2 by considering the surface element in the following approximation form
\begin{eqnarray}
    d\mathcal{S}  = r \left[1-\frac{2 m }{r}+\frac{Q^2}{r^2}-\frac{\mathcal{N}_q}{r^8}\right]^{-\frac{3}{2}} dr d\phi\simeq r\left[1+\frac{3 m}{r}+\frac{15 m^2-3 Q^2}{2 r^2}\right] dr d\phi.\label{ds3}
\end{eqnarray}

For the choice of $\kappa\lambda$ =3/2, the  optical Gaussian curvature (\ref{cur}) becomes
\begin{eqnarray}
\mathcal{G}_{\kappa\lambda = \frac{3}{2}} &=& -\frac{2 m}{r^3}+\frac{3(m^2 +Q^2)}{r^4}-\frac{6 m Q^2}{r^5}+\frac{2 Q^4}{r^6}-\frac{36 \mathcal{N}_q}{r^{10}}+\frac{66 m \mathcal{N}_q}{r^{11}}-\frac{31 \mathcal{N}_q Q^2}{r^{12}}+\frac{20 \mathcal{N}_q^2}{r^{18}}.\label{cur3}
\end{eqnarray}

Similarly, on using Eqns. (\ref{ds3}) and (\ref{cur3}) in Eq. (\ref{DA}) the obtain the deflection angle is obtained as 
\begin{eqnarray}
    \Theta_{\kappa\lambda = \frac{3}{2}} &\simeq& \frac{4 m}{b}+ \frac{3 \pi  \left(m^2-Q^2\right)}{4 b^2}+\Bigg[\frac{315 \pi }{256 b^8}+\frac{512 m}{135 b^9}+\frac{63 \pi  \left(72 m^2-23 Q^2\right)}{2560 b^{10}}-\frac{512 \left(165 m^3-64 m Q^2\right)}{2541 b^{11}}\nonumber
    \\
    &&+\frac{7161 \pi  q \left(5 m^2-Q^4\right)}{8192 b^{12}}\Bigg]\mathcal{N}_q+ \Bigg[\frac{60775 \pi  \left(Q^2-5 m^2\right)}{196608 b^{18}}-\frac{262144 m}{123981 b^{17}}-\frac{32175 \pi }{131072 b^{16}}\Bigg]\mathcal{N}_q^2.
\end{eqnarray}

 It is clear at once that the surrounding field structure parameter $\mathcal{N}_q$ here also affects the deflection angle. Interestingly, the deflection reduces to the deflection angle of Reissner–Nordstrom black for  $\mathcal{N}_q$ = 0 and of the Schwarzchild black hole for $\mathcal{N}_q$ = 0 and $Q$ = 0.
\begin{table*}[thth]
\caption{ Inter and outer horizons $r_{\mp}$, photon sphere radius $r_p$, shadow radius $r_s$, and impact parameter $b(r_s)$ for $\mathcal{N}_{q} = 0.1$.} \label{tab1}
\label{tab2}       
\centering
\begin{tabular}{|c|c|c|c|c|c|c|c|c|c|c|c|c|c|c|c|c|}
\hline
 \multicolumn{12}{|c|}{ $\kappa\lambda$ = -3/2} \\
\hline
 \multicolumn{6}{|c|}{ $m$ = 1} & \multicolumn{6}{|c|}{$Q$ = 0.5}\\
\hline
$Q$ & $r_-$ &   $r_+$       &  $r_p$  &  $r_s$ &   $b(r_s)$ & m & $r_-$ &   $r_+$       &  $r_p$  &  $r_s$ &   $b(r_s)$   \\
\hline
0    & 2.11617 &  2.11617  &  3.17641  & 5.52365 & 7.05821    & 1    & 0.129042  &  1.98906 &  3.00829 & 5.30667 &  6.81867\\
\hline
0.3  & 0.0447999  & 2.07226  &   3.11802 & 5.44793 & 6.97447   & 1.5     & 0.083978  &  3.04276   & 4.58065 & 8.00563 &  10.2482\\
\hline
0.6 & 0.191449  & 1.92722 &   2.92762 & 5.20381 & 6.70562    & 2    & 0.062580  &  4.07101  & 6.11942 & 10.65990 &  13.6274\\
\hline
0.9 & 0.515117 &  1.60537  & 2.5276 &  4.71076 & 6.1696  & 2.5     & 0.049950  & 5.08934  &  7.64488 & 13.296 &  16.9856\\
\hline
\end{tabular}\label{table1}
\end{table*}
\begin{table*}[thth]
\caption{Inter and outer horizons $r_{\mp}$, photon sphere radius $r_p$, shadow radius $r_s$, and impact parameter $b(r_s)$ for $\mathcal{N}_{q} = -0.1$.} \label{tab2}
\label{tab2}       
\centering
\begin{tabular}{|c|c|c|c|c|c|c|c|c|c|c|c|c|c|c|c|c|}
\hline
 \multicolumn{12}{|c|}{ $\kappa\lambda$ = 3/2} \\
\hline
 \multicolumn{6}{|c|}{ $m$ = 1} & \multicolumn{6}{|c|}{$Q$ = 0.5}\\
\hline
$Q$ & $r_-$ &   $r_+$       &  $r_p$  &  $r_s$ &   $b(r_s)$ & $m$ & $r_-$ &   $r_+$       &  $r_p$  &  $r_s$ &    $b(r_s)$   \\
\hline
0    & 0.692653  &  1.99922  &  2.99977 &  5.19603 & 6.62525   & 1    & 0.727302 &  1.86465   & 2.8225 & 4.96772 &  6.37342\\
\hline
0.3  & 0.703873  &  1.95299  &   2.93848 & 5.11666  &  6.53752  & 1.5     & 0.653533 &  2.91416  & 4.38598 & 7.64662 &  9.77492\\
\hline
0.3 & 0.746912 &  1.79815 &  2.73645 & 4.85845 & 6.25351    & 2    & 0.615793  &  3.93648  & 5.91547 & 10.2827 &  13.1295\\
\hline
0.9 & 0.905747  &  1.42177  & 2.29155 & 4.31822 &  5.66871  & 2.5     & 0.590678  &  4.94949  &  7.43273 & 12.9031 &  16.4668\\
\hline
\end{tabular}\label{table1}
\end{table*}

\section{ ANALYSIS of the Deflection angles} \label{sec5}

In this section, we are going to analyze the detailed behaviour of the obtained deflection angles $\Theta_{\kappa\lambda = -\frac{3}{2}}$ and $\Theta_{\kappa\lambda = \frac{3}{2}}$ with respect to the impact parameter $b$, black hole charge $Q$, and black hole mass $m$. Needfully, we   demonstrate both the deflection angles $\Theta_{\kappa\lambda = -\frac{3}{2}}$ and $\Theta_{\kappa\lambda = \frac{3}{2}}$ against the impact parameter $b$ in Figs. \ref{fig5} and \ref{fig6}, respectively by varying charge $Q$ = $\{0, 0.3, 0.6, 0.9\}$ with fixed $m = 1$ and by varying mass $m$ = $\{1, 1.5, 2, 2.5\}$ with fixed $Q$ = 0.5.

 \begin{itemize}

\item {\bf Deflection Angle vs Impact Parameter:} Both  the Figs. \ref{fig5} and \ref{fig6} clearly show that the obtained deflection angles are decreasing in nature for increasing values of impact parameter $b$. Thus, the deflection angles are inversely proportional to the impact parameter $b$.
 
 \item {\bf Deflection Angle vs Black Hole Charge:} One can easily observe from Figs. \ref{fig5} (Left) and \ref{fig6} (left) that the present deflection angles also decrease as black hole charge $Q$ increases. Therefore,   the deflection angles are also inversely proportional to the black hole charge $Q$.
 
\item {\bf Deflection Angle vs Black Hole Mass:}  Figs. \ref{fig5} (Right) and \ref{fig6} (Right) ensure that both the deflection angles increase as black hole mass $m$ increases, and hence,  the deflection angles are directly proportional to the mass of the black hole.

\end{itemize}

In addition, we compare our obtained deflection angles with the deflection angles SC and RN of the Schwarzchild and Reissner–Nordstrom black holes, respectively in Fig. \ref{fig7}. The deflection angles follow the order as follows: RN $\approx$ $\Theta_{\kappa\lambda = \frac{3}{2}} < SC < \Theta_{\kappa\lambda = -\frac{3}{2}}$. Thus, the surrounding quintessence field in the Rastall gravity black hole results in a greater deflection angle than in Schwarzschild or Reissner–Nordström black holes for positive $\mathcal{N}_q$. Conversely, for negative $\mathcal{N}_q$, the deflection angle approaches that of the Reissner–Nordström black hole in this model.

\begin{figure}[!htbp]
\begin{center}
\begin{tabular}{rl}
\includegraphics[width=6.5cm]{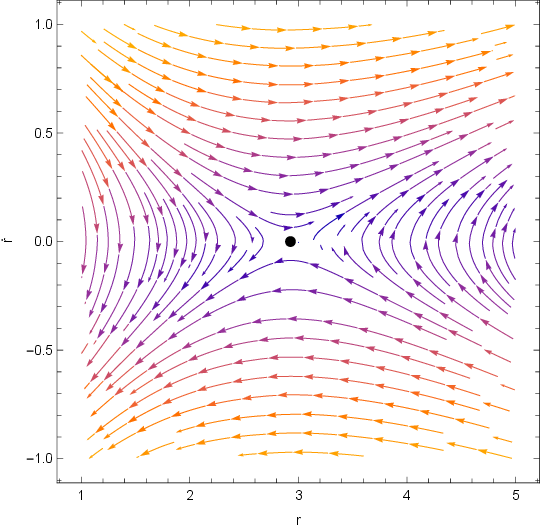}
\end{tabular}
\begin{tabular}{rl}
\includegraphics[width=6.5cm]{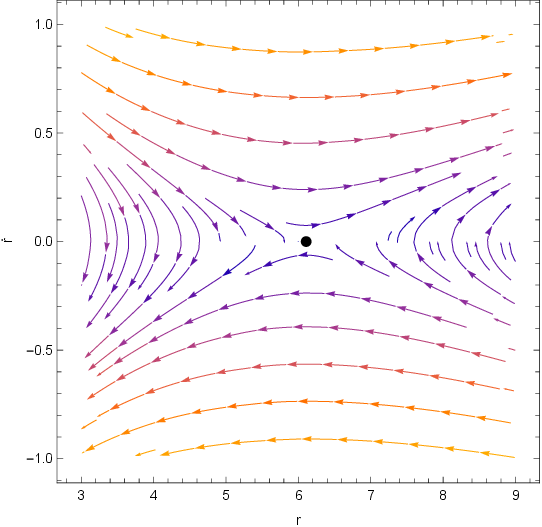}
\end{tabular}
\end{center}
\caption{ Profiles of the trajectories of photon rear the photon sphere corresponding to $\kappa\lambda = -3/2$ along with $\mathcal{N}_q = 0.1$, $m = 1$, $Q$ = 0.6 (Left) and $\kappa\lambda = 3/2$ along with $\mathcal{N}_q = -0.1$, $m = 2$, $Q$ = 0.5 (Right).}\label{fig10}
\end{figure}

\section{Shadow cast of the black hole} \label{sec6}
Here, we are going to study the shadow cast for the present black hole. In this context, we first describe the photon sphere of the black hole.
\subsection{Photon Sphere}

In order to study the null geodesics for the black hole surrounded by quintessence in Rastall gravity (\ref{bh}), we use the Hamilton-Jacobi method. The particle motion around the black hole (\ref{bh}) can be governed by the following Hamilton-Jacobi equation
\begin{eqnarray}
    \frac{\partial \mathcal{J}}{\partial \sigma} =-\frac{1}{2}g^{\alpha\beta}\frac{\partial \mathcal{J}}{\partial x^\alpha} \frac{\partial \mathcal{J}}{\partial x^\beta}.
\end{eqnarray}
 where $\mathcal{J}$ denotes the Jacobi action. Also, the Lagrangian of a test particle in the equatorial plane of the black hole (\ref{bh}) can be defined as 
 \begin{eqnarray}
     \mathcal{L} = \frac{1}{2} \left(g_{tt}\dot{t}^2+g_{rr}\dot{r}^2+g_{\phi\phi}\dot{\phi}^2\right).\label{L}
 \end{eqnarray}

 Now, the separable form of the Jacobi action for the photon is expressed as
\begin{eqnarray}
    \mathcal{J} = -Et+l\phi+\mathcal{J}_r(r)+\mathcal{J}_\theta(\theta),
\end{eqnarray}
where $\mathcal{J}_r(r)$ and $\mathcal{J}_\theta(\theta)$ correspond to the functions of $r$
and $\theta$, respectively. Also, $E$ and $l$ are  energy and angular momentum, which can be given for the metric (\ref{bh}) as
\begin{eqnarray}
    E &=&\frac{d\mathcal{L}}{dt} =  -\mathcal{F}(r)\dot{t},~~~~~~~ l= \frac{d\mathcal{L}}{d\phi} = r^2 \sin^2\theta \dot{\phi}.
\end{eqnarray}
 Thus, the geodesic equations can be expressed as
\begin{eqnarray}
    \frac{dt}{d\sigma} &=& \frac{E}{\mathcal{F}(r)}, ~~
    \frac{d\phi}{d\sigma} = -\frac{l}{r^2 \sin^2\theta},~~
    \frac{dr}{d\sigma}=\pm \frac{\sqrt{\mathcal{R}(r)}}{r^2},~~
    \frac{d\theta}{d\sigma}=\pm \frac{\sqrt{\Theta(\theta)}}{r^2},\label{em}
\end{eqnarray}
where
\begin{eqnarray}
    \mathcal{R}(r) = r^4E^2-(\mathcal{C}+l^2)r^2 \mathcal{F}(r),~~~\Theta(\theta) = \mathcal{C}-l^2 \cot\theta,
\end{eqnarray}
with $\mathcal{C}$ as a constant termed as the Carter separation constant.

\begin{figure}[!htbp]
\begin{center}
\begin{tabular}{rl}
\includegraphics[width=6.5cm]{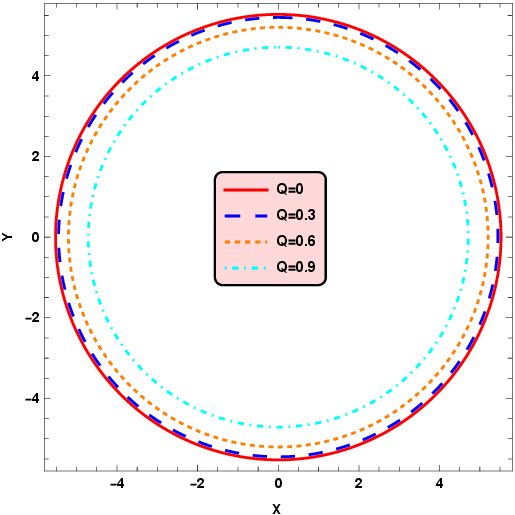}
\end{tabular}
\begin{tabular}{rl}
\includegraphics[width=6.5cm]{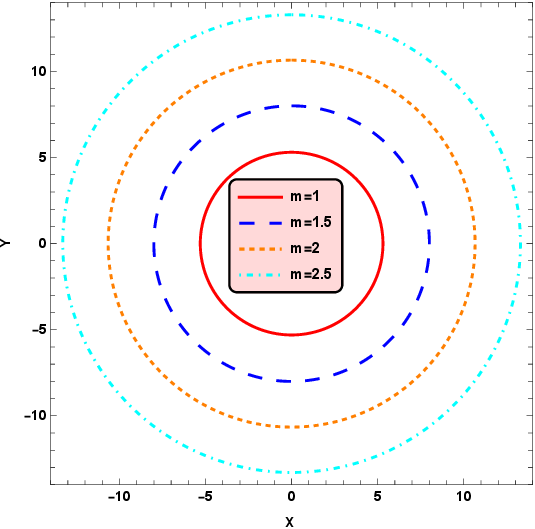}
\end{tabular}
\end{center}
\caption{Profiles of the shadows of the black hole corresponding to $\kappa\lambda = -3/2$ along with $\mathcal{N}_q = 0.1$, $m = 1$ (Left) and $\kappa\lambda = -3/2$ along with $\mathcal{N}_q = 0.1$, $Q = 0.5$ (Right).}\label{fig11}
\end{figure}

Now, we take into account two dimensionless impact parameters $\eta$ and $\xi$ depending on $\mathcal{C}$, $E$, and $l$, defined as
\begin{eqnarray}
    \eta = \frac{\mathcal{C}}{E^2},~~~~\xi = \frac{l}{E}.
\end{eqnarray} 
Thus, the above dimensionless impact parameters generate the solution $\mathcal{R}(r)$ in the following form
\begin{eqnarray}
    \mathcal{R}(r) = r^4E^2-r^2 E^2(\xi^2+\eta)\mathcal{F}(r).\label{R}
\end{eqnarray}
Here, we will focus on the spherical photon orbit of fixed radius $r$ with $\dot{r} = 0$ and $\ddot{r} = 0$. This fixed radius $r = r_{p}$ for the spherical photon orbit is known as the photon sphere radius. For a far observer, the photon comes to the black hole (\ref{bh}) near its equatorial plane and the unstable circular orbits follow
\begin{eqnarray}
\mathcal{R}(r)|_{r=r_{p}} = 0,~\text{and}~ \mathcal{R}'(r)|_{r=r_{p}} = 0.\label{RC}
\end{eqnarray}

Now, the first condition of the above Eq. (\ref{RC}) yields the following result
\begin{eqnarray}
\xi^2+\eta = \frac{r_{p}^2}{\mathcal{F}(r_{p})}.\label{xieta}
\end{eqnarray}
 Also, the simultaneous fulfilment of both conditions in Eq. (\ref{RC}) yields the following result for the described black hole (\ref{bh}) 
\begin{eqnarray}
    r_p^{\frac{3}{1-\kappa\lambda}} \left[2 (\kappa\lambda-1)(3 m-r) r_p^{\frac{\kappa\lambda+2}{\kappa\lambda-1}} +(4 \kappa\lambda-1) \mathcal{N}_q\right]-4 (\kappa\lambda-1) Q^2= 0.\label{ps}
\end{eqnarray}
One can see that the above Eq. (\ref{ps}) has two positive real roots for $\kappa\lambda$ = -3/2, 3/2. We consider the greatest positive real root of the above Eq. (\ref{ps}) as the radius of the photon sphere $r_p$, given in the tables-\ref{tab1} and \ref{tab2}. In both the cases $\kappa\lambda$ = -3/2, 3/2, the photon sphere radius decreases for increasing values of charge $Q$ and increases for increasing values of mass $m$, clear from  tables-\ref{tab1} and \ref{tab2}.

\begin{figure}[!htbp]
\begin{center}
\begin{tabular}{rl}
\includegraphics[width=6.5cm]{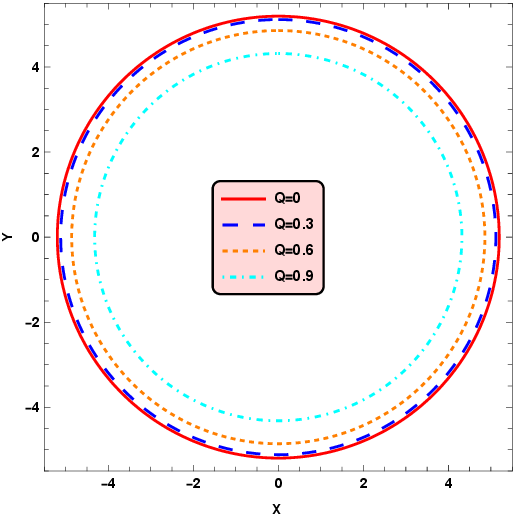}
\end{tabular}
\begin{tabular}{rl}
\includegraphics[width=6.5cm]{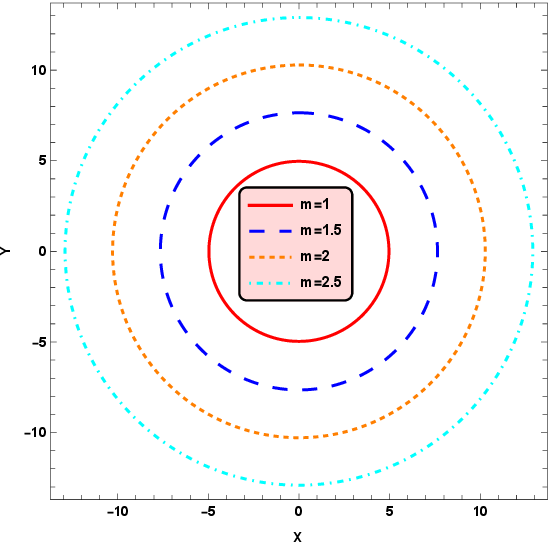}
\end{tabular}
\end{center}
\caption{Profiles of the shadows of the black hole corresponding to $\kappa\lambda = 3/2$ along with $\mathcal{N}_q = -0.1$, $m = 1$ (Left) and  $\kappa\lambda = 3/2$ along with $\mathcal{N}_q = -0.1$, $Q = 0.5$ (Right).}\label{fig12}
\end{figure}

Now, the effective potential of the black hole (\ref{bh}) at the equatorial plane can be defined as \cite{sg42}
\begin{eqnarray}
    V_{\text{eff}}(r)= \frac{\mathcal{F}(r)}{r^2}=\frac{1}{r^2}\left(1-\frac{2 m }{r}+\frac{Q^2}{r^2}-\frac{\mathcal{N}_q}{r^{\frac{-1-2\kappa\lambda}{1-\kappa\lambda}}}\right).
\end{eqnarray} 

The conditions for the photon sphere orbit, $\dot{r} = 0$ and $\ddot{r} = 0$, ensure that the effective potential meets the following requirements
\begin{eqnarray}
    V_{\text{eff}}(r)|_{r=r_p} = \frac{1}{b_p^2},~~~V'_{\text{eff}}(r)|_{r=r_p} = 0.\label{veff}
\end{eqnarray}
where $b$ is the impact parameter defined as $b = |l|/E$. We display the effective potential against the radial coordinate $r$ for $\kappa$ = -3/2 and 3/2 in Figs. \ref{fig8} and \ref{fig9}, respectively. For both cases $\kappa$ = -3/2 and 3/2, the maximum value of the effective potential associated with the photon orbit increases and shifts closer to the black hole singularity as the charge parameter $Q$ increases. In contrast, the maximum value of the effective potential exhibits the opposite behaviour as the black hole mass $m$ increases. Indeed, the maximum effective potential permits an unstable circular orbit around the black hole and a photon can leave this orbit if it is disturbed by external gravitational forces or interactions with other particles. We also demonstrate the photon trajectories near the photon sphere in Fig. \ref{fig10} for $\kappa\lambda = -3/2$, $\mathcal{N}_q = 0.1$, $m = 1$, $Q$ = 0.6 (Left) and $\kappa\lambda = 3/2$, $\mathcal{N}_q = -0.1$, $m = 2$, $Q$ = 0.5 (Right), clear that the photon sphere are unstable.

\subsection{Shadow Radius}
Here, we are going to estimate the shadow radius of the present black hole. In order to do this, we consider the celestial coordinates $X$ and $Y$, defined as \cite{sv04}
\begin{eqnarray}
    X &=& \lim_{r_0\rightarrow \infty}\left(-r_0\sin\theta_0\left[\frac{d\phi}{dr}\right]_{r_0,\theta_0}\right),~~~~
    Y = \lim_{r_0\rightarrow \infty}\left(r_0\left[\frac{d\theta}{dr}\right]_{r_0,\theta_0}\right),\label{y}
\end{eqnarray}
where $(r_0, \theta_0)$ represents the position coordinate of the observer. For the null geodesic, the above coordinates (\ref{y}) can be expressed as
\begin{eqnarray}
    X &=& -\frac{\xi}{\sin\theta},~~~~~ Y = \pm \sqrt{\eta -\xi^2 \cot^2\theta},\label{y1}
\end{eqnarray}

Let us assume that the observer is positioned on the equatorial plane $(\theta = \pi/2)$, then, the celestial coordinates (\ref{y1}) read as
\begin{eqnarray}
    X &=& -\xi, ~~~~~ Y = \pm \sqrt{\eta}.
\end{eqnarray}

\begin{figure}[!htbp]
\begin{center}
\begin{tabular}{rl}
\includegraphics[width=8cm]{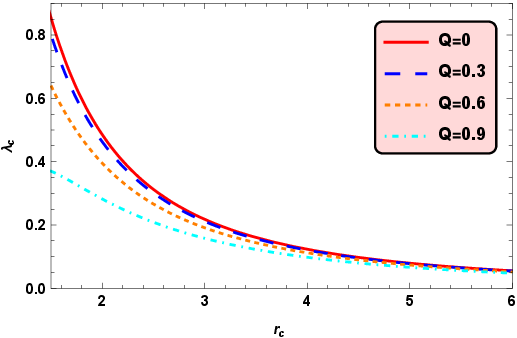}
\end{tabular}
\begin{tabular}{rl}
\includegraphics[width=8cm]{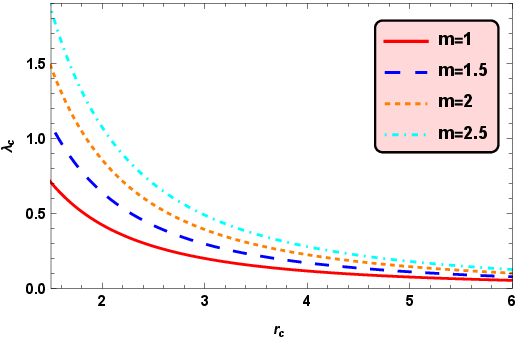}
\end{tabular}
\end{center}
\caption{ Profiles of the coordinate time Lyapunov exponent $\lambda_c$ against the  radial coordinate $r$ corresponding to $\kappa\lambda = -3/2$ along with $\mathcal{N}_q = 0.1$, $m = 1$ (Left) and $\kappa\lambda = -3/2$ along with $\mathcal{N}_q = 0.1$, $Q = 0.5$ (Right).}\label{fig13}
\end{figure}

Thus, on using the result (\ref{xieta}), the shadow radius $r_{s}$ for our described black hole (\ref{bh}) can be expressed as
\begin{eqnarray}
    r_{s} = \sqrt{X^2 + Y^2} = \sqrt{\xi^2 +\eta}= \frac{r_{p}}{\sqrt{\mathcal{F}(r_{p})}} = r_p\left[1-\frac{2 m }{r_p}+\frac{Q^2}{r_p^2}-\frac{\mathcal{N}_q}{r_p^{\frac{-1-2\kappa\lambda}{1-\kappa\lambda}}}\right]^{-\frac{1}{2}}.\label{rs}
\end{eqnarray}
We demonstrate the shadow radius of the present black hole graphically in Figs. \ref{fig11} and \ref{fig12} for  $\kappa$ = -3/2 and 3/2, respectively. In both cases,  the shadow radius decreases with increasing charge parameter $Q$ (See Figs. \ref{fig11} (Left) and \ref{fig12} (Left)) and increases for increasing mass $m$ (See Figs. \ref{fig11} (Right) and \ref{fig12} (Right)).  Moreover, we estimate the shadow radius based on Eq. (\ref{rs}), impact parameter $b(r_s)$ based on Eq. (\ref{veff}), given in tables-\ref{tab1} and \ref{tab2} for $\kappa$ = -3/2 and 3/2, respectively. One can easily see that the shadow radius and impact parameter both exhibit the same behaviour as the photon sphere radius for $Q$ and $m$.

\section{ Quasinormal Modes via  Lyapunov exponents} \label{sec7}
In this section, we are willing to discuss the quasinormal modes (QNMs) with the help of Lyapunov exponents for the present black hole (\ref{bh}). 

\subsection{Lyapunov Exponents}

In a dynamical system, the Lyapunov exponent or Lyapunov characteristic exponent quantifies the average rate expansion and contraction of adjacent trajectories in the phase space. The negative Lyapunov exponent indicates convergence of nearby trajectories while the positive Lyapunov exponent is responsible for the divergence between nearby geodesics, where the system's trajectory is highly sensitive to change the initial conditions. However, the vanishing Lyapunov exponent indicates the presence of marginal stability. For the geodesic stability, the equation of motion in terms of Lyapunov exponents can be written as
\begin{eqnarray}
    \frac{d\mathcal{Y}_i}{dt} = \mathcal{H}_i(\mathcal{Y}_j).\label{Y}
\end{eqnarray}

The linearized form of the above Eq. (\ref{Y}) around a certain orbit can be defined as 
\begin{eqnarray}
    \frac{d\delta\mathcal{Y}_i}{dt} =  \mathcal{A}_{ij}(t)\delta\mathcal{Y}_j(t),\label{l1}
\end{eqnarray}

where $\mathcal{A}_{ij}(t)=\frac{\partial\mathcal{H}_i}{\partial \mathcal{Y}_j}|_{\mathcal{Y}_i(t)}$ denoted the linear stability matrix \cite{nc03}. Now, the solution of above Eq. (\ref{l1}) can be expressed in the following form
\begin{eqnarray}
    \delta\mathcal{Y}_i(t) =  \mathcal{X}_{ij}(t)\delta\mathcal{Y}_j(0),\label{l2}
\end{eqnarray}

where $\mathcal{X}_{ij}(t)$ denoted the evolution matrix satisfying the following relation
\begin{eqnarray}
    \dot{\mathcal{X}}_{ij}(t) = \mathcal{A}_{im} \mathcal{X}_{mj}(t),
\end{eqnarray}
with $\mathcal{X}_{ij}(0) = \partial_{ij}$.
\begin{figure}[!htbp]
\begin{center}
\begin{tabular}{rl}
\includegraphics[width=8cm]{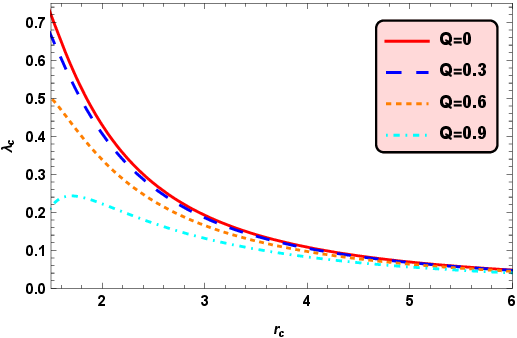}
\end{tabular}
\begin{tabular}{rl}
\includegraphics[width=8cm]{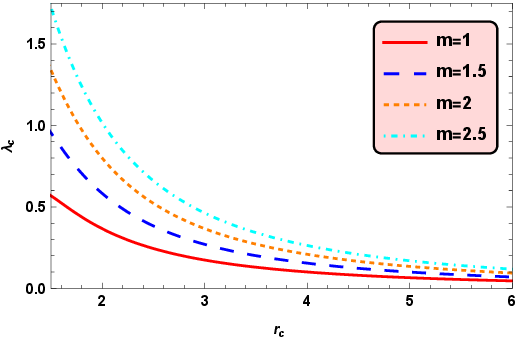}
\end{tabular}
\end{center}
\caption{Profiles of the coordinate time Lyapunov exponent $\lambda_c$ against the  radial coordinate $r$ corresponding to $\kappa\lambda = 3/2$ along with $\mathcal{N}_q = -0.1$, $m = 1$ (Left) and $\kappa\lambda = 3/2$ along with $\mathcal{N}_q = -0.1$, $Q = 0.5$ (Right).}\label{fig14}
\end{figure}

Now, the principal Lyapunov exponent $\lambda$ can be represented with the help of the eigenvalues of $\mathcal{X}_{ij}$ as 
\begin{eqnarray}
    \lambda = \lim_{x \to a} \frac{1}{t}\log \left(\frac{\mathcal{X}_{ij}(t)}{\mathcal{X}_{mj}(0)}\right).
\end{eqnarray}

If there is a set of $n$ Lyapunov exponents associated with an $n$-dimensional
independent system, they can be ordered by magnitude as follows
\begin{eqnarray}
    \lambda_1 \geq \lambda_2 \geq \lambda_3 \geq ........\geq \lambda_n.
\end{eqnarray}

It is important to note that the set of $n$ Lyapunov exponents $\lambda_i$ is known as the Lyapunov spectrum.

Now, the canonical momenta for the Lagrangian (\ref{L}) can be expressed as
\begin{eqnarray}
    P_q = \frac{\partial\mathcal{L}}{\partial\dot{q}}.
\end{eqnarray}

Thus, the generalized momenta can be expressed in the following form
\begin{eqnarray}
    P_t &=& g_{tt} \dot{t} = -E,~~
    P_\phi = g_{\phi \phi} \dot{\phi} = l,~~
    P_r = g_{rr} \dot{r}.
\end{eqnarray}

Now, the Euler-Lagrange equation of motion yields the following result
\begin{eqnarray}
    \frac{dP_q}{d\tau} = \frac{\delta \mathcal{L}}{\delta q}.\label{pq}
\end{eqnarray}
From the above result, we can obtain the non-linear differential equation in two-dimensional phase-space with respect to the phase-space variables $\mathcal{Y}_i(t) =(P_r, r)$ in the following forms
\begin{eqnarray}
    \frac{dP_r}{d\tau} = \frac{\partial \mathcal{L}}{\partial q},~~~\text{and}~~~ \frac{dr}{d\tau} = \frac{P_r}{g_{rr}}.
\end{eqnarray}

Now, the linearization of the above equation of motion  about the circular orbit, where  $r$ is considered to be a constant $r_c$, generates the following  infinitesimal evolutionary matrix 
\begin{eqnarray}
A = \begin{bmatrix}
0 & \frac{d}{dr}\left(\frac{\partial \mathcal{L}}{\partial r}\right) \\
-\frac{1}{g_{rr}} & 0  
\end{bmatrix}_{r=r_c}.\label{mat}
\end{eqnarray}

The eigenvalues of the above matrix are known as the principal Lyapunov exponent which can inform the stability of the orbit. Thus, the eigenvalues of the evolution matrix (\ref{mat}) along circular orbits satisfy the following result 
\begin{eqnarray}
    \lambda^2 = \frac{1}{g_{rr}} \frac{d}{dr}\left(\frac{\partial \mathcal{L}}{\partial r}\right).\label{lam}
\end{eqnarray}

\begin{figure}[!htbp]
\begin{center}
\begin{tabular}{rl}
\includegraphics[width=8cm]{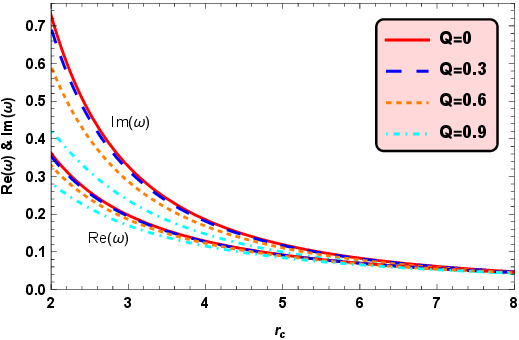}
\end{tabular}
\begin{tabular}{rl}
\includegraphics[width=8cm]{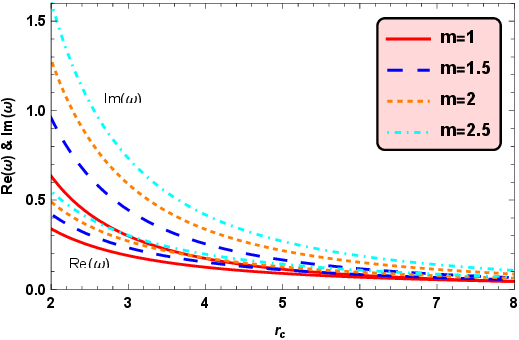}
\end{tabular}

\end{center}
\caption{ Profiles of the real and imaginary parts of the QNMs frequency against $r_c$ corresponding to $\kappa\lambda = -3/2$ along with $\mathcal{N}_q = 0.1$, $m = 1$ (Left) and $\kappa\lambda = -3/2$ along with $\mathcal{N}_q = 0.1$, $Q = 0.5$ (Right).}\label{fig15}
\end{figure}

Again, we can write the Lagrange equation of motion in the following form
\begin{eqnarray}
    \frac{d}{d\tau}\left(\frac{\partial\mathcal{L}}{\partial\dot{r}}\right)-\frac{\partial\mathcal{L}}{\partial r} = 0.
\end{eqnarray}

Thus, the Lyapunov exponent can be expressed in terms of the square of radial velocity $\dot{r}^2$  as
\begin{eqnarray}
   \frac{\partial\mathcal{L}}{\partial r} = \frac{1}{2g_{rr}}\frac{d}{dr}(\dot{r} g_{rr})^2.\label{le}
\end{eqnarray}

Finally, Eqs. (\ref{lam}) and (\ref{le}) simultaneously yield the principal Lyapunov exponent in the following form
\begin{eqnarray}
    \lambda^2 = \frac{1}{2}\frac{1}{g_{rr}}\frac{d}{dr}\left[\frac{1}{g_{rr}}\frac{d}{dr}(\dot{r}g_{rr})^2\right].\label{lam1}
\end{eqnarray}
It is important to note here that the circular geodesics satisfies $\dot{r}^2 = \left(\dot{r}^2\right)' = 0$ \cite{sc83}.

Therefore, we can obtain the proper time Lyapunov exponent $\lambda_p$ and coordinate time Lyapunov exponent $\lambda_c$ from (\ref{lam1}) as \cite{vc09}
\begin{eqnarray}
    \lambda_p = \pm \sqrt{\frac{(\dot{r}^2)^{''}}{2}},\label{lamp}
\end{eqnarray}
\begin{eqnarray}
    \lambda_c = \pm \sqrt{\frac{(\dot{r}^2)^{''}}{2\dot{t}^2}}.\label{lamc}
\end{eqnarray}

It is noted that the Lyapunov exponents occurred in $\pm$ pairs to conserve the volume of the phase space. For further study, we shall consider only the positive Lyapunov exponent. The circular orbit is stable when $\lambda_p$ or $\lambda_c$ is imaginary, while the circular orbit is unstable when $\lambda_p$ or $\lambda_c$  is real. Moreover, the circular orbit becomes marginally stable or a saddle point for $\lambda_p$ = 0 or $\lambda_c$ = 0. The literature ensures that the results (\ref{lamp}) and (\ref{lamc})  are satisfied by the spherically symmetric BH space-times \cite{pp11, dp11, mr11} and axisymmetric space-time \cite{pp13, pp78, pp13a}. 

In this study, we shall only compute the coordinate time Lyapunov exponent $\lambda_c$  for the null circular geodesics, as there is no proper time for photons. For the optical metric (\ref{bh1}) of the present black hole (\ref{bh}), we obtain coordinate time Lyapunov exponent $\lambda_c$ as follows
\begin{eqnarray}
    \lambda_c &=& \sqrt{\left(\frac{(2 \kappa\lambda+1) \mathcal{N}_q r_c^{-\frac{3}{\kappa\lambda-1}}}{2 (\kappa\lambda-1)}+m r_c-Q^2\right) \left(\frac{(4 \kappa\lambda-1) \mathcal{N}_q \left[\kappa\lambda \left(5 r_c^6-3\right)-2 r_c^6+3\right] r_c^{-\frac{3 (4 \kappa\lambda-3)}{\kappa\lambda-1}}}{2(\kappa\lambda-1)^2}+\frac{3 m r_c-4 Q^2}{r_c^6}\right)}.\label{lamcF}
\end{eqnarray}

It is important to note that the above result (\ref{lamcF}) reduces to  the Lyapunov exponent $\lambda^{RN}_c$ = $\sqrt{(mr_c-Q^2)(3mr_c-4Q^2)/r_c^6}$ for Reissner–Nordstrøm black hole, where $\mathcal{N}_q$ = 0 \cite{pp16}. Moreover, the result (\ref{lamcF}) yields the Lyapunov exponent for Schwarzchild black hole $\lambda^{Shh}_c$ = $\sqrt{3}m/r_c^2$, where $\mathcal{N}_q$ = $Q$ = 0 \cite{pp16}. The obtained Lyapunov exponent (\ref{lamcF}) is real in both the cases $\kappa\lambda = -3/2$ and 3/2 (See Figs. \ref{fig13}-\ref{fig14}), which is ready to present the unstable circular orbit. In particular, this result ensures that the photon sphere in this study is unstable.  

\begin{figure}[!htbp]
\begin{center}
\begin{tabular}{rl}
\includegraphics[width=8cm]{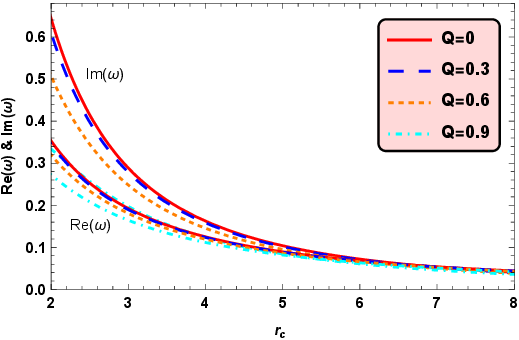}
\end{tabular}
\begin{tabular}{rl}
\includegraphics[width=8cm]{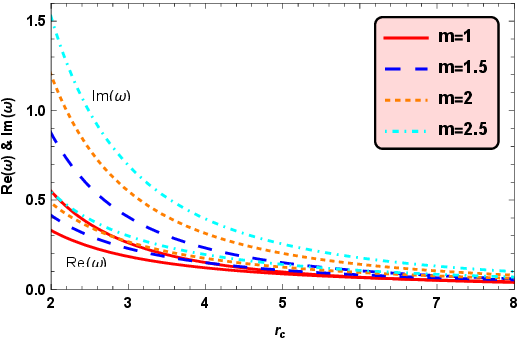}
\end{tabular}

\end{center}
\caption{Profiles of the real and imaginary parts of the QNMs frequency against $r_c$ corresponding to $\kappa\lambda = 3/2$ along with $\mathcal{N}_q = -0.1$, $m = 1$ (Left) and $\kappa\lambda = 3/2$ along with $\mathcal{N}_q = -0.1$, $Q = 0.5$ (Right).}\label{fig16}
\end{figure}

\subsection{Quasinormal Modes Frequency}

To study the QNMs frequency, we consider the usual wave-like equation of the radial part of the perturbation variable $\mathcal{X}$ in the following form \cite{si87}
\begin{eqnarray}
\frac{d^2 \mathcal{X}}{dr_*^2} + \xi \mathcal{X} =0,\label{X}
\end{eqnarray}
where,
\begin{eqnarray}
    \xi = \omega^2 -V(r).
\end{eqnarray}
Here, $r_*$ is the tortoise coordinate given by $r_* = \int \frac{1}{\mathcal{F}(r)} dr$, $\omega$ is the QNMs frequency which we want to figure out, and  $V_{RW}(r)$ is known as the Regge-Wheeler potential. For the present black hole (\ref{bh}) $V_{RW}(r)$ can be expressed as
\begin{eqnarray}
    V_{RW}(r) &=&  \mathcal{F}(r)\left[\frac{j(j+1)}{r^2}+\frac{1-s^2}{r}\frac{\partial F(r)}{\partial r}\right].\label{VV}
\end{eqnarray}
where $j$ is the spherical harmonic index, and  $s$ is the spin of the perturbation, which is $s$ = 0 for the scalar case. 

Thus, the result (\ref{VV}) reads as
\begin{eqnarray}
    V_{RW}(r) &=&  \left(1-\frac{2 m }{r}+\frac{Q^2}{r^2}-\frac{\mathcal{N}_q}{r^{\frac{-1-2\kappa\lambda}{1-\kappa\lambda}}}\right)\left(\frac{j(j+1)}{r^2}+\frac{2 m r (\kappa \lambda -1)+\mathcal{N}_q (2 k \lambda +1) r^{\frac{3}{1-\kappa \lambda }}+Q (2-2 \kappa \lambda )}{r^4 (k \lambda -1)}\right).
\end{eqnarray}
Now, on using the the eikonal limit $j \rightarrow \infty$, we obtain
\begin{eqnarray}
    \xi \approx\omega^2 -\frac{j^2}{r^2} \left(1-\frac{2 m }{r}+\frac{Q^2}{r^2}-\frac{\mathcal{N}_q}{r^{\frac{-1-2\kappa\lambda}{1-\kappa\lambda}}}\right).\label{xi}
\end{eqnarray}
We can find the  maximum value of $\xi$ (\ref{xi}) occurred at $r = r_0$, which implies
\begin{eqnarray}
  (k \lambda -1)( 6 m r_0 -4Q^2 -2r_0^2)+(4 k \lambda-1)  \mathcal{N}_q r_0^{\frac{3}{1-k \lambda }} = 0.\label{max}
\end{eqnarray}

\begin{figure}[!htbp]
\begin{center}
\begin{tabular}{rl}
\includegraphics[width=8cm]{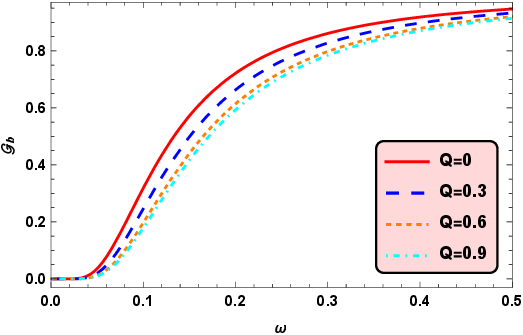}
\end{tabular}
\begin{tabular}{rl}
\includegraphics[width=8cm]{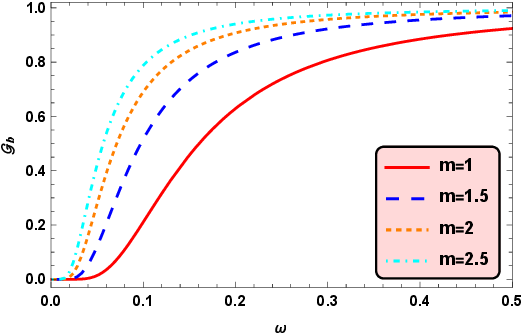}
\end{tabular}
\end{center}
\caption{Profiles of greybody factor $\mathcal{G}_b$ against the  frequency parameter $\omega$ corresponding to $\kappa\lambda = -3/2$ along with $\mathcal{N}_q = 0.1$, $m = 1$ (Left) and $\kappa\lambda = -3/2$ along with $\mathcal{N}_q = 0.1$, $Q = 0.5$ (Right).}\label{fig17}
\end{figure}

Also, the null circular geodesic at $r = r_c$ yields
\begin{eqnarray}
  (k \lambda -1)( 6 m r_c -4Q^2 -2r_c^2)+(4 k \lambda-1)  \mathcal{N}_q r_c^{\frac{3}{1-k \lambda }} =0.\label{max}
\end{eqnarray}

Since the maximum value of $\xi$ coincides with the null circular geodesics at $r_c = r_0$ then we can find the QMNs conditions from Eq. (\ref{X}) as follows \cite{si87, bf85, si87a}
\begin{eqnarray}
    \frac{\xi(r_0)}{\sqrt{-2\left(\frac{d^2\xi}{dr^2_*}\right)_{r=r_0}}} = \frac{i}{2}(2n+1).\label{nn}
\end{eqnarray}
where $n$ represents the overtone number. It is noted that the above equation is estimated at the maximum values of $\xi$ (\ref{xi}) occurring at $r = r_0 = r_c$ i.e. $\frac{d\xi}{dr_*}$ = 0 at $r = r_c$.

Thus, for the large limit of $j$ the above equation (\ref{nn}) defines the QNMs frequency $\omega$ as \cite{vc09}
\begin{eqnarray}
    \omega_{QNM}&=&j \Omega_c -\frac{i}{2}(2n+1) \lambda_c,
\end{eqnarray}
where $\Omega_c$ is the angular frequency measured by the asymptotic observers defined as $\Omega_c =\dot{\phi}/\dot{t} = \sqrt{F'(r_c)/2r_c}$.

Therefore, we obtain the QNMs frequency $\omega$ for the black hole (\ref{bh}) as
\begin{eqnarray}
    \omega&=&j \sqrt{\frac{(2 \kappa\lambda+1) \mathcal{N}_q r_c^{-\frac{4 \kappa\lambda-1}{\kappa\lambda-1}}}{2 (\kappa\lambda-1)}+\frac{m r_c-Q^2}{r_c^4}}-\frac{i}{2}(2n+1) \sqrt{\left(\frac{(2 \kappa\lambda+1) \mathcal{N}_q r_c^{-\frac{3}{\kappa\lambda-1}}}{2 (\kappa\lambda-1)}+m r_c-Q^2\right)}\times\nonumber
    \\
    &&\sqrt{\frac{(4 \kappa\lambda-1) \mathcal{N}_q \left[\kappa\lambda \left(5 r_c^6-3\right)-2 r_c^6+3\right] r_c^{-\frac{3 (4 \kappa\lambda-3)}{\kappa\lambda-1}}}{2(\kappa\lambda-1)^2}+\frac{3mr_c-4 Q^2}{r_c^6}}.\label{QNM}
\end{eqnarray}

It is worth mentioning that the above result (\ref{QNM}) is crucial in the eikonal limit, where the real and imaginary parts of the quasinormal modes of the black hole (\ref{bh}) are given by the frequency and the instability timescale of the unstable null circular geodesics. Here, we demonstrate the real and imaginary parts of $\omega$ in Figs. \ref{fig15}-\ref{fig16}, these show that both the real and imaginary parts of $\omega$ decrease for increasing $Q$ and increase for increasing $m$ in both the cases $\kappa\lambda$ = -3/2 and 3/2. Notably,  Gogoi et al. \cite{go21} investigated the QNMs of black holes in Rastall gravity in the presence of quintessence field using the  5th order WKB approximation method. Their results showed that the real quasinormal frequencies increase with increasing $Q$. In the case of imaginary quasinormal frequencies, there is a small dip in the variation pattern, followed by an increase again. 

For Reissner–Nordstrøm black hole ($\mathcal{N}_q$ = 0), the above result (\ref{QNM}) yields
\begin{eqnarray}
    \omega^{RN}&=&j \sqrt{\frac{m r_c-Q^2}{r_c^4}}-\frac{i}{2}(2n+1) \sqrt{\frac{\left(m r_c-Q^2\right)(3 m r_c-4 Q^2)}{r_c^6}}.\label{QNM1}
\end{eqnarray}

For Schwarzchild black hole ($\mathcal{N}_q$ = Q = 0), the result (\ref{QNM}) gives
\begin{eqnarray}
    \omega^{Sch}&=&j \sqrt{\frac{m}{r_c^3}}-\frac{i}{2}(2n+1) \frac{\sqrt{3} m}{r_c^2}.\label{QNM1}
\end{eqnarray}

\begin{figure}[!htbp]
\begin{center}
\begin{tabular}{rl}
\includegraphics[width=8cm]{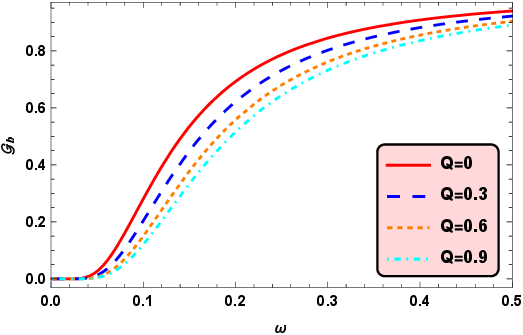}
\end{tabular}
\begin{tabular}{rl}
\includegraphics[width=8cm]{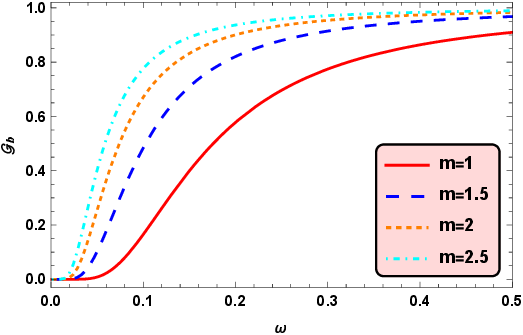}
\end{tabular}
\end{center}
\caption{Profiles of greybody factor $\mathcal{G}_b$ against the  frequency parameter $\omega$ corresponding to $\kappa\lambda = 3/2$ along with $\mathcal{N}_q = -0.1$, $m = 1$ (Left) and $\kappa\lambda = 3/2$ along with $\mathcal{N}_q = -0.1$, $Q = 0.5$ (Right).}\label{fig18}
\end{figure}

Interestingly, our obtained QNMs frequencies for Reissner–Nordström and Schwarzschild black holes are matched with the results given in Ref. \cite{pp16}.

\section{GREYBODY factor of the black hole} \label{sec8}

Here, we discuss the lower bound of the greybody factor (GBF) for the reported black hole (\ref{bh}). The GBF describes the emissivity of the black hole and it has a nice relation to the quantum nature of the black hole. The lower bounds of the GBF can be written as \cite{mv99}
\begin{eqnarray}
    \mathcal{G}_b \geq \text{sech}^2\left(\frac{1}{2\omega}\int_{-\infty}^{\infty} V(r)dr^*\right).\label{T}
\end{eqnarray}
Now, we shall use the Regge-Wheeler potential given in Eq. (\ref{VV}) to calculate the GBF of the massless scalar. Thus, the lower bound value of GBF (\ref{T}) reads as 
\begin{eqnarray}
    \mathcal{G}_b \geq \text{sech}^2\left(\frac{1}{2\omega}\int_{r_+}^{\infty} \left[ \frac{1}{r}\partial_r g_{rr}+j(j+1)\frac{1}{r^2}\right]dr\right), \label{T2}
\end{eqnarray}
where $r_+$ is the outer event horizon of the black hole. For black hole (\ref{bh}), we obtain the lower bound of GBF as follows
\begin{eqnarray}
    \mathcal{G}_b \geq \text{sech}^2\left(\frac{1}{6 k \lambda  r_+^3 \omega }\left[\mathcal{N}_q (2 k \lambda +1) r_+^{\frac{3}{1-k \lambda }}+k \lambda  \left\{3 r_+ (j(j+1) r_++m)-2 Q^2\right\}\right]\right).\label{T2}
\end{eqnarray}
The lower bound of GBF is demonstrated against the QNMs frequency $\omega$ in Figs. \ref{fig17} and \ref{fig18} by taking $j = 0$ and the respective outer horizon $r_+$ given in Tables-\ref{tab1} and \ref{tab2}. From Figs. \ref{fig17} and \ref{fig18}, one can see that the estimated lower bound is decreasing for increasing charge $Q$, and hence, the lower bound follows the inverse relationship with $Q$. In contrast, the lower bound of GBF increases with the increasing values of mass $m$ i.e. the high mass of the black hole generates the higher values of the GBF. 

For Reissner–Nordstrøm black hole ($\mathcal{N}_q$ = 0) with $r_+=m+\sqrt{m^2-Q^2}$, the above result (\ref{T2}) yields
\begin{eqnarray}
    \mathcal{G}^{RN}_b &\geq& \text{sech}^2\left(\frac{1}{6 \omega r_+^3}\left[3 r_+ (j(j+1) r_++m)-2 Q^2\right]\right),\nonumber
    \\
    &=& \text{sech}^2\left(\frac{3 \left(m+\sqrt{m^2-Q^2}\right) \left[l (l+1) \left(m+\sqrt{m^2-Q^2}\right)+m\right]-2 Q^2}{6 \omega  \left(m+\sqrt{m^2-Q^2}\right)^3}\right).\label{T3}
\end{eqnarray}
For Schwarzchild black hole ($\mathcal{N}_q$ = Q = 0) with $r_+=2m$, the result (\ref{T2}) gives
\begin{eqnarray}
    \mathcal{G}^{Sch}_b \geq \text{sech}^2\left(\frac{1}{2 \omega r_+^2}\left[j(j+1) r_++m\right]\right) = \text{sech}^2\left(\frac{1}{8m \omega}\left[2j(j+1)+1\right]\right).\label{T4}
\end{eqnarray}
One can check that the above results (\ref{T3}) and (\ref{T4}) matched for Reissner–Nordström black hole \cite{ac20} and  Schwarzchild black hole \cite{pb08}.

\section{Results and conclusion} \label{sec9}

In this article, we have studied some thermodynamic features, deflection angle in the weak-field approximation, shadow, quasinormal modes using Lyapunov exponents, and the lower bound of the greybody factor for a charged black hole surrounded by a quintessence field in Rastall gravity. The present study is performed for the  two cases: {\bf (i)} $\kappa\lambda$ = -3/2 with $\mathcal{N}_q$ = 0.1, and {\bf (ii)} $\kappa\lambda$ = 3/2 with $\mathcal{N}_q$ = -0.1. It is worth mentioning that both cases nicely respect the asymptotic nature of the black hole and weak energy condition. In both cases, the present black hole has both the inner and outer event horizons except the case $\kappa\lambda = -3/2$, $\mathcal{N}_q = 0.1$ with $Q = 0$, clear from  Fig. \ref{fig1}. The inner horizons increase for increasing values of charge $Q$ and decrease for increasing values of mass $m$, whereas the outer horizons decrease for increasing values of charge $Q$ and increase for increasing values of mass $m$, this behaviour is also confirmed by the mass profile in Fig. \ref{fig2}. However, for $\kappa\lambda = -3/2$, $\mathcal{N}_q = 0.1$ with $Q = 0$, the black hole exhibits one event horizon, and hence, the uncharged Kiselev-like black hole surrounded by quintessence in Rastal gravity has one event horizon like the Schwarzschild black hole. The heat capacity $C(r_+)$ in Fig. \ref{fig3} ensures that the present black hole solutions are always stable, in which the gravitational redshift $Z(r)$ is highly increasing as it approaches the singularity of the black hole, clear from Fig. \ref{fig4}. All these significant results in both our considered cases have enriched for further study. In order to estimate the deflection angle $\Theta$,  we have considered the propagation of photons on the equatorial plane that yields the optical metrics and Gaussian optical curvatures for the present black holes. The obtained deflection angles  $\Theta_{\kappa\lambda = -\frac{3}{2}}$ and $\Theta_{\kappa\lambda = \frac{3}{2}}$ are analyzed against the impact parameter $b$, black hole charge $Q$, and mass $m$ in Figs. \ref{fig5} and \ref{fig6}, respectively. The deflection angles exhibit the following behaviours:

\begin{itemize}
\item The deflection angles $\Theta_{\kappa\lambda = -\frac{3}{2}}$ and $\Theta_{\kappa\lambda = \frac{3}{2}}$ both are inversely related to the impact parameter $b$ i.e. they reduce for increasing $b$, clear Figs. \ref{fig5} and \ref{fig6}.  
 \item  The deflection angle $\Theta_{\kappa\lambda = -\frac{3}{2}}$ reduces for increasing values of $Q$ (See Fig. \ref{fig5} (Left)) and increases for increasing values of mass $m$ (See Fig. \ref{fig5} (Right)). The deflection angle $\Theta_{\kappa\lambda = \frac{3}{2}}$ exhibits the same characteristics as $\Theta_{\kappa\lambda = -\frac{3}{2}}$ against the black hole charge $Q$ and mass $m$ (See Fig. \ref{fig6}).
\end{itemize}

It is important to note that the obtained deflection angles reduce to $4m/b$+ $3\pi\left(m^2-Q^2\right)/4 b^2$, the deflection angle of the Reissner–Nordstrom black hole \cite{yk22, gy22} for $\mathcal{N}_q$ = 0, and $4m/b$ + $3\pi m^2/4 b^2$, the deflection angle of the Schwarzchild black hole \cite{yk22, jb03} for $\mathcal{N}_q$ = 0 and $Q$ = 0. To check the effect of the quintessence field on the deflection angles, we have compared the deflection angles with the deflection angles of  Schwarzchild black hole $SC$ and Reissner–Nordstrom black hole (RN) in Fig. \ref{fig7}, found that the deflection angles follow: RN $\approx$ $\Theta_{\kappa\lambda = \frac{3}{2}} < SC < \Theta_{\kappa\lambda = -\frac{3}{2}}$. Therefore, the quintessence field surrounding the black hole in Rastall gravity produces a higher deflection angle than Schwarzschild or Reissner–Nordström black holes with positive $\mathcal{N}_q$. Also, RN $<$ SC confirms that the presence of charge $Q$ can reduce the deflection angle. To study the shadow of the present black hole, we have first discussed the photon sphere which is the unstable circular orbit around the black hole. The unstableness of the photon sphere is confirmed by the profiles of the effective potential in Figs. \ref{fig8} and \ref{fig9}, in which the effective potential becomes maximum at the radius of the photon sphere. In addition, we have demonstrated the trajectories of photons near the photon sphere in Fig. \ref{fig10}, which confirms that the photon sphere is an unstable orbit. The shadow radius for the present black hole decreases with increasing charge parameter $Q$ (See Figs. \ref{fig11} (Left) and \ref{fig12} (Left)) and increases for increasing mass $m$ (See Figs. \ref{fig11} (Right) and \ref{fig12} (Right)). The numerical values of the inner and outer event horizons $r_{\mp}$, the radius of the photon sphere $r_p$, the shadow radius $r_s$, and the impact parameter $b(r_s)$ for $\kappa\lambda$ = -3/2 and 3/2 are provided in Tables-\ref{tab1} and \ref{tab2}, respectively. Interestingly, all the numerical values respect the above mentioned results against $Q$ and $m$. We have also analyzed the QNMs for the considered black holes. In this context, we have first derived the coordinate time Lyapunov exponent $\lambda_c$, which is real for both the cases $\kappa\lambda = -3/2$ and 3/2 (See Figs. \ref{fig13}-\ref{fig14}). Thus, the real values of $\lambda_c$ ensure the unstable behaviour of the photon sphere. Moreover, it has been seen that $\lambda_c$ reduces to $\lambda^{RN}_c$ = $\sqrt{(mr_c-Q^2)(3mr_c-4Q^2)/r_c^6}$ for Reissner–Nordstrøm black hole, where $\mathcal{N}_q$ = 0 \cite{pp16} and $\lambda^{Shh}_c$ = $\sqrt{3}m/r_c^2$, where $\mathcal{N}_q$ = $Q$ = 0 \cite{pp16}. The obtained QNMs frequency $\omega$ plays a significant role in the eikonal limit, where the real and imaginary parts of the quasinormal modes of the reported black holes are given by the frequency and the instability timescale of the unstable null circular geodesics. We have also analyzed the real and imaginary parts of $\omega$ graphically in Figs. \ref{fig15}-\ref{fig16} for $\kappa\lambda$ = -3/2 and 3/2, respectively. The real as well as imaginary parts of $\omega$ are decreasing for increasing $Q$ and increasing for increasing $m$ in both cases. At last, we have derived the lower bound of the GBF $\mathcal{G}_b$. Indeed, the  GBF provides insights into the emission properties of black holes by quantifying the probability that particles or radiation originating near the black hole will escape its gravitational pull and reach an external observer. There is a bound on the greybody factor that sets the limit on the minimum transmission probability for particles or radiation to escape a black hole’s gravitational field. The estimated GBF $\mathcal{G}_b$ is analyzed with respect to the quasinormal modes frequency $\omega$ in Figs. \ref{fig17} and \ref{fig18} by taking $j = 0$, indicating that the lower bound of $\mathcal{G}_b$ decreases with increasing black hole charge $Q$ and increases with increasing values of mass $m$. Therefore, the lower bound maintains the inverse relationship with $Q$ and the proportional relationship with $m$.

In conclusion, the present study offers valuable insights for future observations that could reveal more detailed effects of parameters $\kappa\lambda$ and $\mathcal{N}_q$ on the deflection angle, shadow, quasinormal modes, and greybody factor of the black hole surrounded by quintessence in the framework of Rastall gravity. Additionally, this work may encourage the scientific community to determine the strong deflection angle for this black hole model.

\section*{Acknowledgement}
FR and TM would like to thank the authorities of the Inter-University Centre for Astronomy and Astrophysics, Pune, India for providing research facilities. FR is also thankful to SERB, DST  $\&$   DST FIST programme  (SR/FST/MS-II/2021/101(C))  for financial support. We are grateful to the reviewer(s) for their constructive and valuable suggestions.

\end{document}